\documentclass[aip,jcp,reprint,twocolumn,superscriptaddress,showpacs,floatfix]{revtex4-2}
\usepackage{amsmath}
\usepackage{graphicx}% Include figure files
\usepackage{dcolumn}% Align table columns on decimal point
\usepackage{bm}% bold math
\usepackage{verbatim}
\usepackage[utf8]{inputenc}
\usepackage[T1]{fontenc}
\usepackage{mathptmx}
\usepackage{etoolbox}
\usepackage{subcaption} % For subfigure support
\usepackage{float}
\usepackage{bm}% bold math
\usepackage{mathrsfs}
\usepackage{multirow}
\usepackage{ragged2e}
\usepackage{placeins}
\DeclareUnicodeCharacter{2212}{-}
\usepackage{xr}
\usepackage{xr-hyper}
\usepackage{hyperref}
\hypersetup{
    colorlinks=true,    % Enables colored links
    linkcolor=blue,     % Color of internal links
    citecolor=magenta,      % Color of citations
    pdfborder={0 0 0}   % Removes the red boxes around links
}
\makeatletter
\def\@email#1#2{%
 \endgroup
 \patchcmd{\titleblock@produce}
  {\frontmatter@RRAPformat}
  {\frontmatter@RRAPformat{\produce@RRAP{*#1\href{mailto:#2}{#2}}}\frontmatter@RRAPformat}
  {}{}
}%
\makeatother

\begin{document}
%\preprint{AIP/123-QED}
\title{A reduced-cost third-order algebraic diagrammatic construction method based on state-specific frozen natural orbitals: Application to the electron-attachment problem}

\author{Tamoghna Mukhopadhyay}
\affiliation{\small Department of Chemistry, Indian Institute of Technology Bombay, Powai, Mumbai 400076, India}
\author{Kamal Majee}
\affiliation{\small Department of Chemistry, Indian Institute of Technology Bombay, Powai, Mumbai 400076, India}
\author{Achintya Kumar Dutta}
\thanks{Corresponding author}
\email[e-mail: ]{achintya@chem.iitb.ac.in}
\affiliation{\small Department of Chemistry, 
Indian Institute of Technology Bombay, Powai, Mumbai 400076, India}
\affiliation{ \small Department of Inorganic Chemistry, Faculty of Natural Sciences, Comenius University, Ilkovičova 6, Mlynská dolina 84215 Bratislava, Slovakia \\
}%
\email[e-mail: ]{achintya.kumar.dutta@uniba.sk}
%\maketitle

\begin{abstract}

We have developed a reduced-cost non-Dyson third-order algebraic diagrammatic construction theory for the electron-attachment problem based on state-specific frozen natural orbitals. Density fitting and truncated natural auxiliary functions were employed to enhance computational efficiency. The use of state-specific frozen natural orbitals significantly decreases the virtual space and provides a notable speedup over the conventional EA-ADC(3) method with a systematically controllable accuracy. A perturbative correction for the truncated natural orbitals significantly reduces the error in the calculated electron affinity values. The method also shows sufficient accuracy in the case of non-valence correlation-bound anions, where the local approximation-based methods fail. The efficiency of the method is demonstrated by performing an EA-ADC(3) calculation with more than 1300 basis functions.

\end{abstract}

\maketitle
%%%%%%%%%%%%%%%%%%%%%%%%%%%%%%%%%%%%%%%%%%%%%%%% Introduction %%%%%%%%%%%%%%%%%%%%%%%%%%%%%%%%%%%%%%%%%%%%%%%%%%%%%%%%
\section{Introduction}
\label{sec1}

Electron attachment to atoms and molecules is a fundamental process in physics, chemistry, and biology, with relevance to phenomena ranging from electron transfer in photosynthesis~\cite{pshenichnyukInterconnectionsDissociativeElectron2018,rubinElectronTransportBiological1984} to radiation-induced damage in nucleic acids~\cite{narayanansjSecondaryElectronAttachmentInduced2023}. Accurate simulation of electron-attachment-induced phenomena is crucial for a comprehensive understanding of these processes. The theoretical methods available for the simulation of electron affinity can be broadly classified into two distinct categories. The first category of methods consists of the so-called $\Delta$-based methods, in which the electron affinity is defined as the difference between the total electronic energies of the anionic and neutral systems, both calculated at the same level of theory. The second class of methods comprises direct energy-difference approaches, which determine the electron affinity as the transition energy between the neutral and anionic systems. The latter methods are particularly advantageous, as they require only a single calculation, provide access to transition probabilities, and circumvent the numerical instabilities often encountered in $\Delta$-based approaches.~\cite{vooraTheoreticalApproachesTreating2017} Among the various direct energy difference-based methods available, the equation of motion coupled cluster (EOM-CC) approach is particularly popular due to its systematically improvable hierarchy~\cite{roweEquationsofMotionMethodExtended1968,nooijenEquationMotionCoupled1995,nooijenCoupledClusterApproach1992,stantonEquationMotionCoupledcluster1993,krylovEquationofMotionCoupledClusterMethods2008}. The EOM-CC method for the electron affinity problem is generally used in the singles and doubles approximation (EA-EOM-CCSD)~\cite{nooijenEquationMotionCoupled1995}. It has a formal scaling of $\mathscr{O}(\mathscr{N}^6)$ and storage requirements that scales as $\mathscr{O}(\mathscr{N}^4)$, where $\mathscr{N}$ is the number of basis functions. The non-Hermitian form of the coupled-cluster similarity-transformed Hamiltonian makes property calculations in the EOM-CCSD method roughly twice as costly as the corresponding energy calculation. ~\cite{stantonEquationMotionCoupledcluster1993}

Algebraic diagrammatic construction (ADC) theory~\cite{schirmerRandomphaseApproximationNew1982,schirmerNewApproachOneparticle1983} provides a Hermitian and size-consistent alternative to the EOM-CC method, with a natural hierarchy defined by perturbation order.  ADC theory was originally developed in the context of Green's function propagator theory.~\cite{schirmerRandomphaseApproximationNew1982} Perturbative expansions of the one-particle Green’s function, commonly referred to as the electron propagator, have led to a variety of computational methods formulated within the framework of the Dyson equation.~\cite{vonniessenComputationalMethodsOneparticle1984,ortizElectronPropagatorPicture1997}. This group of methods is commonly referred to as Dyson-ADC methods. Later, Schirmer et al.~\cite{schirmerNewApproachOneparticle1983} developed the non-Dyson formulation of ADC, in which the electron-attachment problem can be solved separately from the ionization potential problem. The ADC equations can also be derived~\cite{banerjeeThirdorderAlgebraicDiagrammatic2019} using the  effective Liouvillian formalism, introduced by Mukherjee and Kutzelnigg~\cite{mukherjeeEffectiveLiouvilleanFormalism1989}. However, the ADC method is generally formulated in the intermediate state representation (ISR)~\cite{schirmerClosedformIntermediateRepresentations1991,mertinsAlgebraicPropagatorApproaches1996,schirmerIntermediateStateRepresentation2004,dreuwAlgebraicDiagrammaticConstruction2023}. In addition to the vertical electron-attachment energy, the $(N+1)$-electronic state wave function is also accessible within the ISR formalism. Dempwolf et al.~\cite{dempwolffIntermediateStateRepresentation2021} have reported property calculations within the EA-ADC intermediate state formalism. 
The second-order ADC method (EA-ADC(2)) often provides insufficient accuracy for the electron attachment problem~\cite{schirmerNonDysonThirdorderApproximation1998,banerjeeThirdorderAlgebraicDiagrammatic2019,tripathiBoundAnionicStates2019}, and the third-order approximation, EA-ADC(3), is required for higher accuracy. Similar to the EA-EOM-CCSD method, the EA-ADC(3) method scales as  $\mathscr{O}(\mathscr{N}^6)$, ), but it is computationally more favorable due to the non-iterative nature of the  $\mathscr{O}(\mathscr{N}^6)$ scaling terms and its lower pre-factor.~\cite{dempwolffEfficientImplementationNonDyson2019} Dreuw and his coworkers have recently explored a fourth-order approximation to the EA-ADC theory, denoted as EA-ADC(4).~\cite{leitnerFourthOrderAlgebraicDiagrammatic2024}  \\
The ADC(3) method cannot be routinely applied to systems containing more than 10–15 atoms without additional approximations. Various strategies have been described in the literature to reduce the computational cost of wave-function-based methods. These strategies involve approximating the two-electron integrals using the density-fitting approximation ~\cite {hattigAdvancesQuantumChemistry2005,banerjeeEfficientImplementationSinglereference2021,mukhopadhyayStatespecificFrozenNatural2023,mannaReducedCostEquation2025,mannaEfficientStateSpecificNatural2024} or employing local  ~\cite{pulayLocalizabilityDynamicElectron1983,pulayOrbitalinvariantFormulationSecondorder1986} and/or natural orbitals.~\cite{lowdinQuantumTheoryManyParticle1955} Among the various flavors  of natural orbitals available~\cite{mesterReducedcostSecondorderAlgebraicdiagrammatic2018,duttaDomainbasedLocalPair2019,haldarEfficientFockSpace2021,duttaPairNaturalOrbital2016,landauFrozenNaturalOrbitals2010,pokhilkoExtensionFrozenNatural2020,helmichPairNaturalOrbital2013,frankPairNaturalOrbital2018,mesterReducedcostLinearresponseCC22017}, the frozen natural orbital~\cite{barrNatureConfigurationInteractionMethod1970} have emerged as the most popular approaches. FNO-based implementation of IP~\cite{mukhopadhyayStatespecificFrozenNatural2023}, DIP~\cite{mandalThirdOrderRelativisticAlgebraic2025}, and EE-ADC~\cite{mesterReducedcostSecondorderAlgebraicdiagrammatic2018,mesterReducedCostSecondOrderAlgebraicDiagrammatic2023} method have been described in the literature. However, to the best of our knowledge, no natural-orbital-based low-cost ADC method has been reported for the electron-attachment problem. 
The aim of this work is to develop a low-cost ADC(3) method for the electron attachment problem based on state-specific frozen natural orbitals. 

\section{Theory}
\label{sec2}
\subsection{Algebraic Diagrammatic Construction (ADC) Theory}
\label{sec:adc}
In the  ISR formalism of ADC, ~\cite{schirmerClosedformIntermediateRepresentations1991,mertinsAlgebraicPropagatorApproaches1996,schirmerIntermediateStateRepresentation2004,dreuwAlgebraicDiagrammaticConstruction2023} electron-attached states are generated by applying a linear operator to the correlated $N$-electron ground-state wave function. 
\begin{equation}
\label{eq1}
    \left|\psi _{A}^{N+1}\right\rangle ={{\hat{C}}_{A}}\left|\psi _{0}^{N}\right\rangle
\end{equation}

The linear operator $\hat{C}_{A}$ can be represented in the second quantized notation as

\begin{equation}
\label{eq2}
    \left\{{{\hat{C}}_{A}}\right\}=\left\{{{\hat{c}}_{a}^{\dagger }},\hat{c}_{b}^{\dagger }\hat{c}_{a}^{\dagger }{{\hat{c}}_{i}},\hat{c}_{c}^{\dagger }\hat{c}_{b}^{\dagger }\hat{c}_{a}^{\dagger }{{\hat{c}}_{i}}{{\hat{c}}_{j}},...; \quad i<j...,a<b<c...\right\}
\end{equation}
where $i, j, k, ...$denote occupied orbitals and  $a, b, c, ...$ denote virtual orbitals. 

These correlated target states are generally non-orthonormal and must therefore be orthonormalized. First, precursor states are constructed through Gram–Schmidt orthogonalization. These states are then transformed into excitation-class-orthonormalized (ECO) ADC intermediate states by symmetric normalization. 
 The ADC shifted Hamiltonian $\left(\hat{H} -E_{0}\right)$ can be expressed as a secular matrix $\left(\mathbf{M}_{AB}\right)$ in the basis of these intermediate states
\begin{equation}
\label{eq3}
    {{\mathbf{M}}_{AB}}=\langle \tilde{\Psi }_{A}^{N+1}|\hat{H}-E_{0}|\tilde{\Psi }_{B}^{N+1}\rangle
\end{equation}
and the exact $(N+1)$-electronic state can be written as
\begin{equation}
\label{eq4}
    \left| \Psi _{b}^{N+1} \right\rangle =\sum\limits_{A}{{{\mathbf{Y}}_{Ab}}}\left| \tilde{\Psi }_{A}^{N+1} \right\rangle
\end{equation}

The ADC equation can be written as the eigenvalue problem 
\begin{equation}
\label{eq6}
    \mathbf{M}\mathbf{Y}=\mathbf{Y} \Omega
\end{equation}
where the eigenvalues $\Omega$ are vertical electron-attachment energies and the eigenvectors $\mathbf{Y}$ determine the spectral amplitudes. The spectral amplitudes $\mathbf{x}$ are obtained from the eigenvector $\mathbf{Y}$ as
\begin{equation}
\label{eq7}
    \mathbf{x}={{\mathbf{Y}}^{\dagger }}\mathbf{f}
\end{equation}
where
\begin{equation}
\label{eq8}
    {{\mathbf{f}}_{Ap}}=\langle \tilde{\psi }_{A}^{N+1}|{{\hat{c}}_{p}^{\dagger }}|\psi _{0}^{N}\rangle
\end{equation}
The secular matrix in Eq.~\eqref{eq3} can be expanded in perturbation order, and truncation at order $n$ leads to ADC($n$) equations
\begin{equation}
\label{eq5}
    \mathbf{M}={{\mathbf{M}}^{(0)}}+{{\mathbf{M}}^{(1)}}+{{\mathbf{M}}^{(2)}}+{{\mathbf{M}}^{(3)}}+...
\end{equation}
Truncating at $n$=2 yields ADC(2) method, whereas truncation at $n$=3 yields ADC(3).  Dreuw and co-workers~\cite{bauerExploringAccuracyUsefulness2022} have proposed a scaled matrix (sm) “fractional-order” ADC scheme, where the third-order contribution to the ADC matrix  is  scaled with an empirical factor x
\begin{equation}
    \mathbf{M}_{sm-ADC[(2)+x(3)]}={{\mathbf{M}}^{(0)}}+{{\mathbf{M}}^{(1)}}+{{\mathbf{M}}^{(2)}}+x{{\mathbf{M}}^{(3)}}
\end{equation}
The scaling parameter $x$ can vary from 0 to 1; the value 0 will lead to the ADC(2) method, and the value 1 will lead to the ADC(3) method. The recommended value of x is 0.5\cite{chakrabortyRelativisticThirdorderAlgebraic2025}.
The $\mathbf{M}$ matrix is generally diagonalized using Davidson's iterative diagonalization method~\cite{davidsonIterativeCalculationFew1975}, which involves the contraction of suitably chosen trial vectors with the Hamiltonian matrix elements to construct the so-called "sigma" vectors.  The programmable expressions for EA-ADC(2) and EA-ADC(3) sigma vectors are provided in the Supplementary Material.  The storage and manipulation of terms involving two-electron integrals can be computationally expensive and requires large storage, especially in extended basis sets. One needs to use an additional approximation to reduce the storage requirements. 
\subsection{Density Fitting (DF) Approximation}
\label{sec:df}
The four-centered two-electron integrals $(pq|rs)$  are defined as ~\cite{hohensteinDensityFittingCholesky2010}
\begin{equation}
\label{eq18}
    (pq|rs)=\int{d{{r}_{1}}\int{d{{r}_{2}}}{{\phi }_{p}}({{r}_{1}}){{\phi }_{q}}({{r}_{1}})\frac{1}{{{r}_{12}}}{{\phi }_{r}}({{r}_{2}}){{\phi }_{s}}({{r}_{2}})}
\end{equation}
The product of orbital functions  ${{\phi }_{x}}(r){{\phi }_{y}}(r)$ can be interpreted as an orbital pair density  ${{\bm{\rho}}_{xy}}$
\begin{equation}
\label{eq19}
    (pq|rs)=\int{d{{r}_{1}}\int{d{{r}_{2}}}{{\bm\rho }_{pq}}({{r}_{1}})\frac{1}{{{r}_{12}}}{{\bm\rho }_{rs}}({{r}_{2}})}
\end{equation}
The pair density ${{\bm\rho }_{xy}}$ can be expanded in an auxiliary basis as 
\begin{equation}
\label{eq20}
    {{\bar{\bm\rho }}_{xy}}(r)=\sum\limits_{P}^{{{N}_{aux}}}{\mathbf{d}_{P}^{xy}{{\chi }_{P}}(r)}
\end{equation}
Here, $\mathbf{d}_{P}^{xy}$denotes the fitting coefficients and $\mathbf{\chi}_{P}$ denotes the auxiliary basis functions. The fitting coefficients are determined by minimizing the Coulomb metric functional,
\begin{equation}
\label{eq21}
    {{\bm\Delta }_{xy}}=\int{d{{r}_{1}}}\int{d{{r}_{2}}}\frac{[{{\bm\rho }_{xy}}({{r}_{1}})-{{{\bar{\bm\rho }}}_{xy}}({{r}_{1}})][{{\bm\rho }_{xy}}({{r}_{2}})-{{{\bar{\bm\rho }}}_{xy}}({{r}_{2}})]}{{{r}_{12}}}
\end{equation}
which leads to
\begin{equation}
\label{eq22}
    \mathbf{d}_{P}^{xy}=\sum\limits_{Q}{(xy|Q){{[{{\mathbf{X}}^{-1}}]}_{QP}}}
\end{equation}
Here, $(xy|Q)$is the three-centered two-electron integral, which can be expressed as 
\begin{equation}
\label{eq23}
    (xy|Q)=\int{d{{r}_{1}}}\int{d{{r}_{2}}{{\phi }_{x}}({{r}_{1}})}{{\phi }_{y}}({{r}_{1}})\frac{1}{{{r}_{12}}}{{\chi }_{Q}}({{r}_{2}})
\end{equation}
where
\begin{equation}
\label{eq24}
    {{\mathbf{X}}_{QP}}=\int{d{{r}_{1}}}\int{d{{r}_{2}}{{\chi }_{Q}}({{r}_{1}})\frac{1}{{{r}_{12}}}}{{\chi }_{P}}({{r}_{2}})
\end{equation}
Therefore, the four-centered two-electron integral $(pq|rs)$ can be expressed in terms of three-centered integrals as ~\cite{dunlapApproximationsApplicationsXa2008,whittenCoulombicPotentialEnergy2003,boysAutomaticFundamentalCalculations1956},
\begin{align}
\label{eq25}
    (pq|rs)&=\int{d{{r}_{1}}\int{d{{r}_{2}}}\sum\limits_{Q}{d_{Q}^{pq}}{{\chi }_{Q}}({{r}_{1}})\frac{1}{{{r}_{12}}}{{\chi }_{r}}({{r}_{2}})}{{\chi }_{s}}({{r}_{2}}) \nonumber \\ 
     & =\sum\limits_{Q}{d_{Q}^{pq}}(Q|rs) \nonumber \\ 
     & =\sum\limits_{PQ}{(pq|P){{[{{\mathbf{X}}^{-1}}]}_{PQ}}(Q|rs)} \nonumber \\ 
     & =\sum\limits_{PQR}{(pq|P){{[{{\mathbf{X}}^{-\frac{1}{2}}}]}_{PQ}}{{[{{\mathbf{X}}^{-\frac{1}{2}}}]}_{QR}}(R|rs)} \nonumber \\ 
     & =\sum\limits_{Q}{\left\{ \sum\limits_{P}{(pq|P){{[{{\mathbf{X}}^{-\frac{1}{2}}}]}_{PQ}}} \right\}\left\{ \sum\limits_{R}{{{[{{\mathbf{X}}^{-\frac{1}{2}}}]}_{QR}}(R|rs)} \right\}} \nonumber \\ 
     & =\sum\limits_{Q}{\mathbf{J}_{pq}^{Q}\mathbf{J}_{rs}^{Q}}  
\end{align}
where
\begin{equation}
\label{eq26}
    \mathbf{J}_{pq}^{Q}={{\sum\limits_{P}{(pq|P)[{{\mathbf{X}}^{-\frac{1}{2}}}]}}_{PQ}}
\end{equation}
The three-centered integrals can be converted into a molecular orbital basis as
\begin{equation}
\label{eq26}
    \mathbf{J}_{mn}^{Q}= {\sum\limits_{pq}{C_{mq}\mathbf{J}_{pq}^{Q}C_{nq}}}
\end{equation}

Subsequently, molecular orbital integrals can be constructed directly from the three-center integrals. In the present implementation, integrals up to two virtual indices are generated and stored, whereas integrals with three and four virtual orbitals are constructed on the fly.  
\subsection{State-Specific Frozen Natural Orbitals (SS-FNO)}
\label{sec:ssfno}
Canonical virtual orbitals are generally not compact, and truncating them often leads to non-systematic errors in the correlation energy. A common approach is to transform the virtual orbital space into a natural orbital basis, which provides systematic convergence of the correlation energy with respect to the size of the virtual space.  Natural orbitals are defined as the eigenfunctions of a correlated one-particle reduced density matrix~\cite{lowdinQuantumTheoryManyParticle1955}.  Among the various types of natural orbitals, \cite{landauFrozenNaturalOrbitals2010,pokhilkoExtensionFrozenNatural2020,barrNatureConfigurationInteractionMethod1970,sosaSelectionReducedVirtual1989,taubeFrozenNaturalOrbitals2005}, the frozen natural orbitals (FNO) are the most popular.~\cite{barrNatureConfigurationInteractionMethod1970} In the FNO approximation, the occupied orbitals are kept frozen at the SCF level, while the virtual space is transformed to the natural orbital basis.  Natural orbitals are obtained by diagonalizing the one-particle reduced density matrix derived from a correlated calculation. 
\begin{equation}
\label{eq9}
    \mathbf{D}\mathbf{V}=\mathbf{V}\eta
\end{equation}
where $D$ is the virtual-virtual block of the one-particle density matrix. 
The eigenvectors ($V$)  are natural orbitals, and the corresponding eigenvalues ($\eta$) give occupation numbers of the natural orbitals. The virtual orbital with a small occupation number generally makes a very small contribution to the ground state correlation energy and the natural orbital with occupation below a critical threshold ($\eta_{crit}$) can be truncated as  
\begin{equation}
\label{eq10}
    \tilde{\mathbf{V}}=\mathbf{V}\mathbf{T}
\end{equation}
Here, tilde ($\sim $) denotes the truncated natural orbital basis and $\mathbf{T}_{ab}$ can be expressed as
\begin{align}
\label{eq11}
    {{\mathbf{T}}_{ab}}&={{\delta }_{ab}} \quad \text{if  } {{\eta}_{a}}>{{\eta}_{crit}} \nonumber \\
              &=0 \quad \quad \text{otherwise}
\end{align}

The virtual-virtual block of the Fock matrix $\mathbf{F}$ is transformed to the truncated natural orbital basis as
\begin{equation}
\label{eq12}
    \tilde{\mathbf{F}} = \tilde{\mathbf{V}}^{\dagger} \mathbf{F} \tilde{\mathbf{V}}
\end{equation}
followed by a diagonalization of the transformed Fock matrix as
\begin{equation}
\label{eq13}
    \tilde{\mathbf{F}}\tilde{\mathbf{Z}}=\tilde{\mathbf{Z}}\tilde{\epsilon }
\end{equation}

to obtain semi-canonical orbitals. The matrix 
\begin{equation}
\label{eq14}
    \mathbf{B}=\tilde{\mathbf{V}}\tilde{\mathbf{Z}}
\end{equation}
connects the canonical virtual molecular orbital basis to the frozen natural virtual orbital basis. The transformation from the atomic orbital basis to the frozen natural orbital basis can be written as 
\begin{equation}
\label{eq15}
    {{\tilde{\mathbf{U}}}_{occ}}={{\mathbf{U}}_{occ}}
\end{equation}
\begin{equation}
\label{eq16}
    {{\tilde{\mathbf{U}}}_{vir}}={{\mathbf{U}}_{vir}}\tilde{\mathbf{V}}\tilde{\mathbf{Z}}={{\mathbf{U}}_{vir}}\mathbf{B}
\end{equation}
where ${{\mathbf{U}}_{occ}}$ and ${{\mathbf{U}}_{vir}}$ are the occupied and virtual blocks of the coefficient matrix for the transformation of atomic orbitals to the canonical molecular orbitals.
The second-order Møller-Plesset perturbation theory (MP2) method~\cite{mollerNoteApproximationTreatment1934} is generally used for the generation of natural orbitals.  However, the reduced density obtained from the correlated ground-state wave function does not contain sufficient information about correlation in the electron-attached state. Consequently, truncation in the standard FNO basis may not yield systematic convergence of electron affinity values. To properly describe the electron-attached state, a first-order approximation to the electron-attached wave function should be used to generate natural orbitals. Consequently, we have used the EA-ADC(2) wave function to generate the natural orbitals. The virtual-virtual block of the one-particle reduced density matrix for the $k^{\text{th}}$ electron-attached state in the ADC(2) method has been defined as
\begin{equation}
\label{eq17}
    \mathbf{D}_{ab}^{SS}(k)=\mathbf{D}_{ab}^{MP2}+\mathbf{D}_{ab}^{EA-ADC(2)}(k)
\end{equation}
where $\mathbf{D}_{ab}^{MP2}$ and $\mathbf{D}_{ab}^{EA-ADC(2)}(k)$ denote the one-particle reduced density matrices corresponding to the ground and the $k^{\text{th}}$ electron-attached state calculated at the MP2 and EA-ADC(2) level of theory, respectively. The explicit expression for $\mathbf{D}_{ab}^{EA-ADC(2)}(k)$ is provided in the Supplementary Material. Natural orbitals are generated from the $\mathbf{D}_{ab}^{SS}(k)$ following the equations (\ref{eq9}-\ref{eq16}). The natural orbitals generated in this procedure will necessarily be state-specific. To distinguish it from the standard MP2-based FNO, it will be denoted as SS-FNO in the rest of the manuscript.  It should be noted that the one-particle reduced density matrix used for generating natural orbitals is constructed using the zeroth-order intermediate-state representation derived from the EA-ADC(2) eigenvectors.

\subsection{Natural Auxiliary Functions (NAF)}
\label{sec:naf}
Analogous to the reduction of the orbital basis using frozen natural orbitals (FNOs), the dimensionality of the auxiliary basis can be reduced using natural auxiliary functions (NAFs).~\cite{kallaySystematicWayCost2014}  NAFs are generated using a singular value decomposition of the three-center two-electron matrix $\mathbf{J}$ in the molecular orbital basis.
\begin{equation}
\label{eq27}
    \mathbf{J}=\mathbf{A}\bm\Sigma {{\mathbf{B}}^{T}}
\end{equation}
Here $\bm\Sigma $ denotes the diagonal matrix containing singular values and $\mathbf{A}$ and $\mathbf{B}$ denote the left and right singular vectors, respectively. Alternatively, the NAFs can be obtained by diagonalizing the matrix 
\begin{equation}
\label{eq28}
    \mathbf{W}=\mathbf{J}{{\mathbf{J}}^{T}}
\end{equation}
Following the recommendation of K\'allay and co-workers~\cite{mesterReducedcostLinearresponseCC22017}, the virtual–virtual block of the three-centered integral matrix  $\mathbf{J}_{ab}$ is used to construct $\mathbf{W}$. NAFs corresponding to eigenvalues smaller than a chosen threshold are discarded. The remaining eigenvectors are collected in the matrix $\tilde{\mathbf{N}}$. The matrix $\mathbf{J}$ is truncated by transforming its auxiliary index as 
\begin{equation}
\label{eq29}
    \tilde{\mathbf{J}}=\mathbf{J}\tilde{\mathbf{N}}
\end{equation}
The  $\tilde{\mathbf{J}}$ are subsequently used in the generation of molecular integrals. 
\subsection{Correction for the truncation}
\label{sec:pc}
Second-order perturbative energy correction has been shown to improve the accuracy of truncated natural orbital-based wave-function methods.~\cite{kallaySystematicWayCost2014,mukhopadhyayStatespecificFrozenNatural2023,mannaEfficientStateSpecificNatural2024} In this study, we have included a correction for the natural orbital-based truncation to the SS-FNO-EA-ADC(3) results. 
\begin{align}
\label{eq30}
    \omega _{SS-FNO-EA-ADC(3)}^{corrected} = &\omega _{SS-FNO-EA-ADC(3)}^{uncorrected}\\
    &+{{\omega }_{canonical-EA-ADC(2)}} \nonumber\\
    &-{{\omega }_{SS-FNO-EA-ADC(2)}} \nonumber
\end{align}

Here, the difference between the EA-ADC(2) results in the canonical and SS-FNO bases is added to the uncorrected EA-ADC(3) result as a correction for both natural orbital and auxiliary basis truncations.

%%%%%%%%%%%%%%%%%%%%%%%%%%%%%%%%%%%%%%%%%%%% Computational Details %%%%%%%%%%%%%%%%%%%%%%%%%%%%%%%%%%%%%%%%%%%%%%%%%%%%
\section{Computational Details}
\label{sec3}
\begin{figure}[h!]
    \centering
    \includegraphics[width=0.45\textwidth]{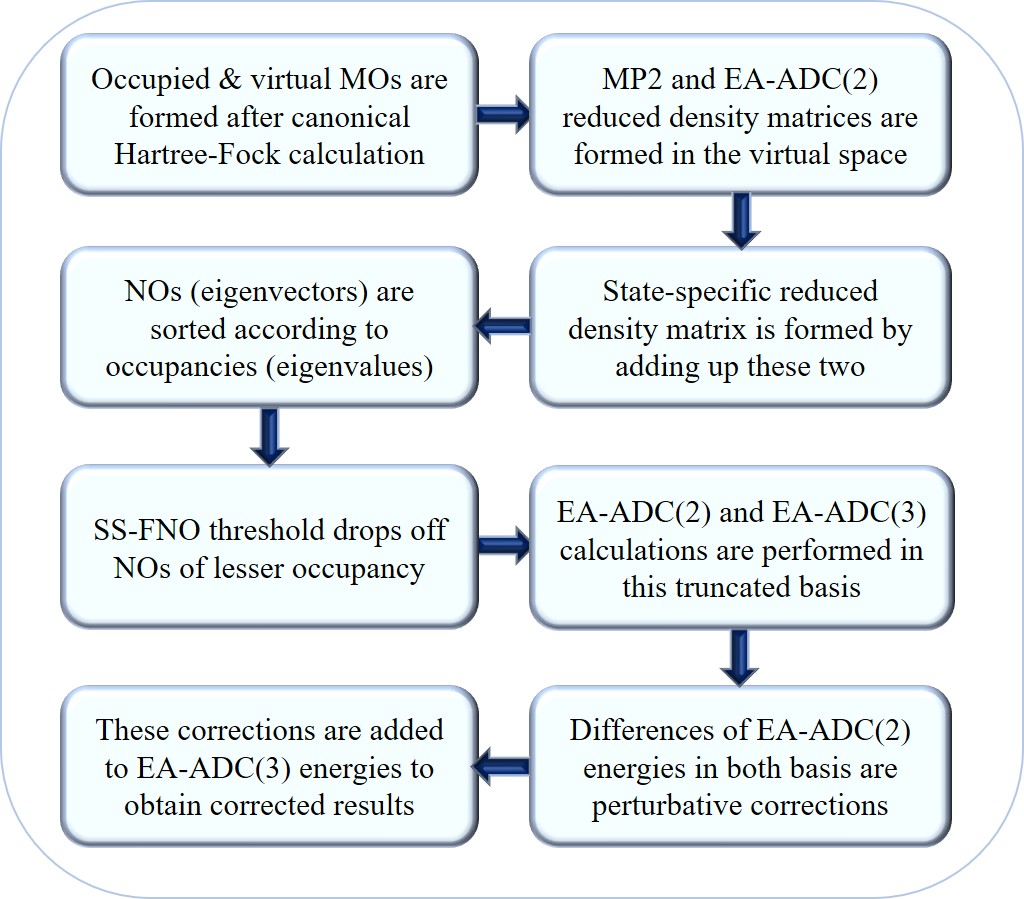}
    \caption{\justifying{The schematic diagram of the algorithm of the SS-FNO-EA-ADC(3) method.}}
    \label{fig:algorithm}
\end{figure}

% BAGH information
The SS-FNO-EA-ADC(3) method is implemented in BAGH,~\cite{duttaBAGHQuantumChemistry2025}, our in-house quantum chemistry software package.  BAGH is primarily written in Python, with computationally intensive components optimized using Cython and Fortran. BAGH is currently integrated with four external packages for the generation of integrals, namely, PySCF~\cite{sunLibcintEfficientGeneral2015,sunPySCFPythonbasedSimulations2018,sunRecentDevelopmentsPySCF2020}, GAMESS-US~\cite{barcaRecentDevelopmentsGeneral2020}, DIRAC~\cite{bastDIRACProgramAtomic2023} and socutils~\cite{wangXubwaSocutils2025}. The implementation of SS-FNO-EA-ADC(3) described in this work uses the PySCF interface for integral generation.
% Algorithm
The steps involved in the SS-FNO-EA-ADC(3) method are as follows:
\begin{enumerate}
    \item After a successful SCF convergence, the three-centered two-electron integrals $(P|ij)$, $(P|ab)$ and $(P|ia)$ are generated.
    \item An MP2 calculation is performed in the canonical basis, and the one-particle reduced density $\left(\mathbf{D}_{ab}^{MP2}\right)$ is constructed.
    \item An EA-ADC(2) calculation is performed in the canonical basis.
    \item The following steps are performed for each root:
    \begin{enumerate}
        \item The virtual-virtual block of the EA-ADC(2) reduced density $\left(\mathbf{D}_{ab}^{EA-ADC(2)}\right)$ is constructed for the corresponding root.
        \item The EA-ADC(2) reduced density matrix is added to the MP2 reduced density matrix to form the state-specific density matrix.
        \item The state-specific density matrix is diagonalized, and the resulting natural orbitals are sorted according to their occupation numbers.
        \item Virtual natural orbitals with occupation numbers below the SS-FNO threshold are discarded.
        \item EA-ADC(2) and EA-ADC(3) calculations are then performed in the truncated natural orbital basis and the corresponding truncated NAF basis.
        \item A perturbative correction, defined as the difference between the canonical and truncated EA-ADC(2) results, is added to the EA-ADC(3) energy.
    \end{enumerate}
\end{enumerate}
A schematic diagram of the SS-FNO-EA-ADC(3) algorithm is shown in FIG.~\ref{fig:algorithm}.

All calculations in this work are performed using the density-fitting approximation. Core electrons were kept frozen for all calculations.

For complete basis set (CBS) extrapolation of electron affinities, the two-point cubic extrapolation scheme introduced by Helgaker and co-workers was employed.~\cite{helgakerBasissetConvergenceCorrelated1997,halkierBasissetConvergenceCorrelated1998}
\begin{equation}
\label{eqn:CBS}
E_{CE}^{x} = E_{CE}^{CBS} + \frac{\alpha}{x^3}
\end{equation}
where $\alpha$ is a fitting parameter and $E_{CE}^{CBS}$ is the correlation energy at the CBS limit. Here, CE stands for correlation energy. The electron affinity results of aug-cc-pVXZ (X=D,T) were extrapolated to the CBS limit using Eq.~\eqref{eqn:CBS}.

%%%%%%%%%%%%%%%%%%%%%%%%%%%%%%%%%%%%%%%%%%%% Results and Discussion %%%%%%%%%%%%%%%%%%%%%%%%%%%%%%%%%%%%%%%%%%%%%%%%%%%
\section{Results and Discussion}
\label{sec4}
\subsection{Optimization of thresholds}
\label{sec:thresh}
To assess the performance of the SS-FNO scheme for the electron-attachment problem, we examined the deviation in EA-ADC(3) electron affinities as a function of the virtual space size in the canonical molecular orbital (MO) basis as well as in the natural orbital bases (FNO and SS-FNO) for the ozone molecule (see Fig.~\ref{fig:pct}). The experimental geometry of the ozone molecule \cite{johnsonComputationalChemistryComparison2002} was used for the calculations. The aug-cc-pVQZ basis set and the aug-cc-pVQZ/C auxiliary basis set were used for the calculations. No NAF-based truncation was used in these calculations. The electron affinities (in eV) computed using different percentages of retained virtual orbitals are listed in the Supplementary Material (Table S1).

\begin{widetext}
\FloatBarrier
%% Threshpct
\begin{figure*}[h!]
\centering
    % First row
    \begin{subfigure}{0.45\textwidth}
        \includegraphics[width=\linewidth]{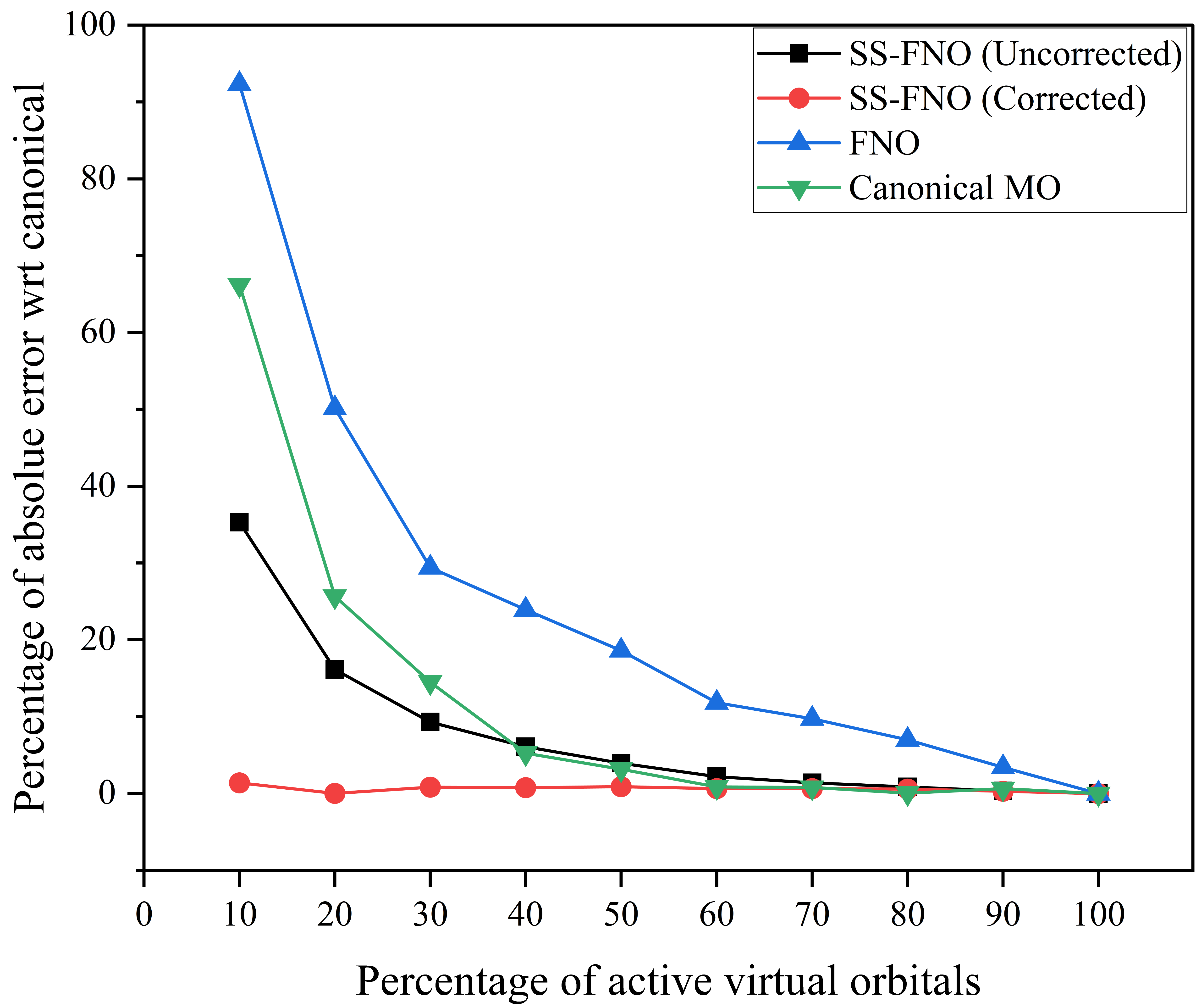}
        \caption{\label{fig:pct}}
    \end{subfigure}
    \hspace{0.025\textwidth}
    \begin{subfigure}{0.45\textwidth}
        \includegraphics[width=\linewidth]{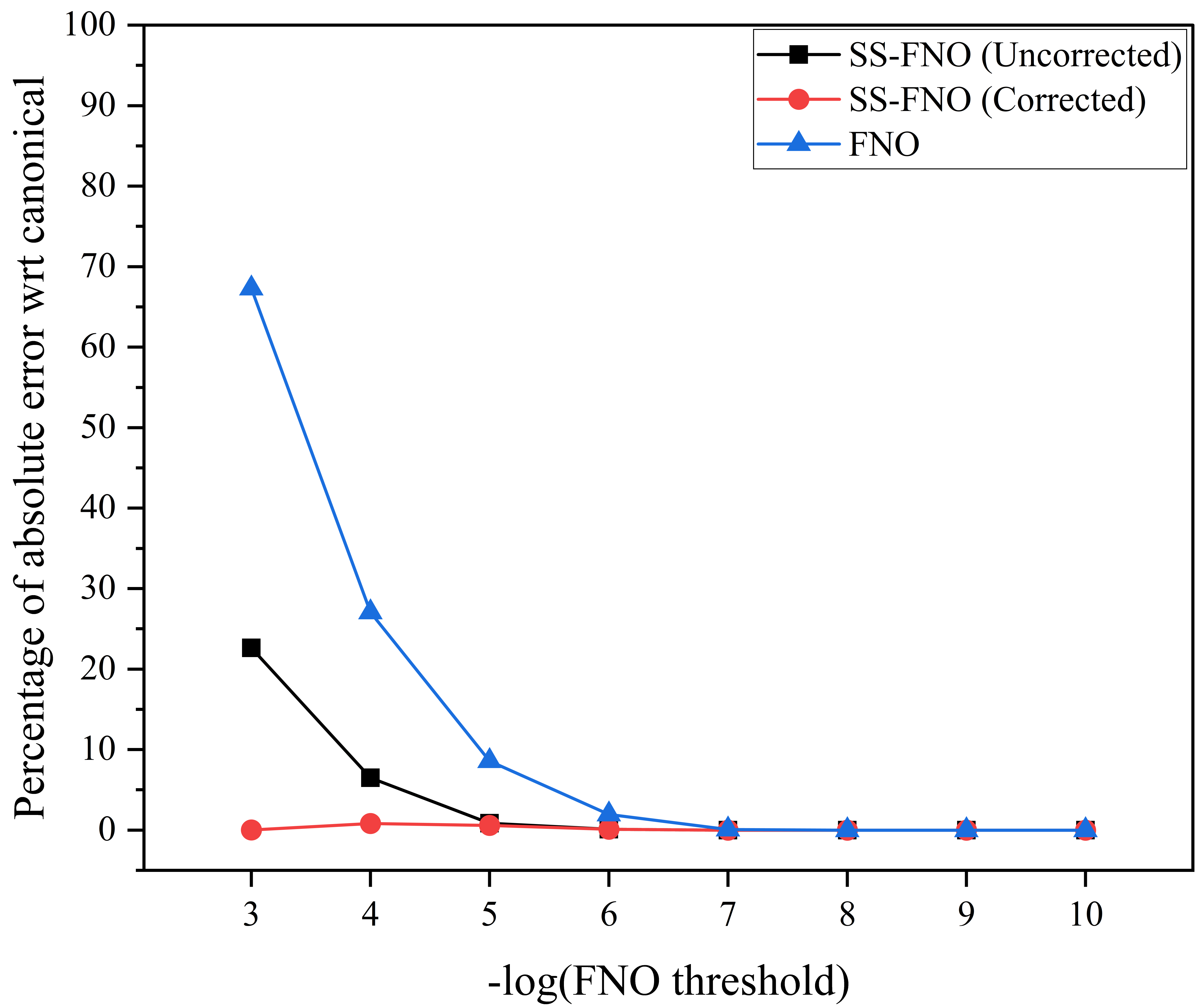}
        \caption{\label{fig:thresh}}
    \end{subfigure}

    % Main caption for the figure
    \caption{
    \label{fig:threshpct}
    \justifying{The comparison of the percentage of absolute errors of EA values (in eV) of O$_3$ molecule in aug-cc-pVQZ basis set for the FNO and SS-FNO versions of EA-ADC(3) with respect to their respective canonical analogues (a) across the percentage of active virtual orbitals and (b) across different truncation thresholds.}
}
\end{figure*}
\FloatBarrier
\end{widetext}
Truncation in the canonical MO basis leads to convergence of the EA values when approximately 60$\%$ of the virtual orbitals are retained. The convergence of electron affinities with respect to the size of the virtual space is even slower in the FNO basis, and the results do not converge even when 90$\%$ of the virtual orbitals are retained. This indicates that natural orbitals derived solely from the ground-state one-particle reduced density matrix do not adequately describe electron-attached states. Unlike the conventional FNO-ADC approach, SS-FNO natural orbitals generated from the EA-ADC(2) one-particle reduced density matrix provide an accurate description of the $(N+1)$-electron state.  It can be seen that, at lower percentages of retained virtual orbitals, the SS-FNO scheme yields smaller errors in EA values compared with canonical truncation. The inclusion of the perturbative correction leads to a significant improvement in the results. The error converges when approximately 30$\%$ of the virtual orbitals are retained. This indicates that the perturbative correction has a substantial impact on the calculated EA values in the truncated SS-FNO basis.

% Nafthresh
\begin{figure}[htbp]
    \centering
    \includegraphics[width=0.45\textwidth]{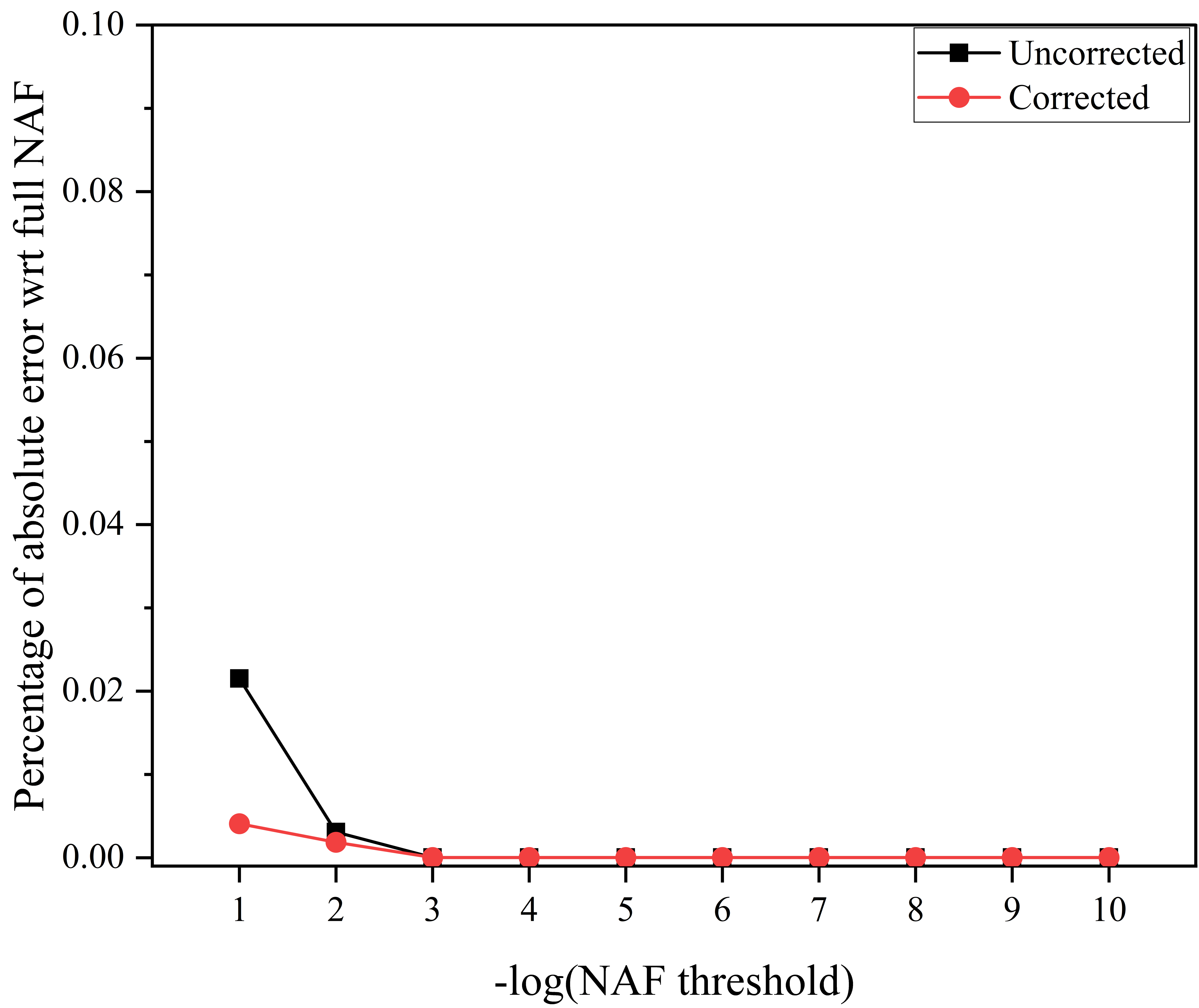}
    \caption{\justifying{The convergence of error (in eV) in SS-FNO-EA-ADC(3) results in aug-cc-pVQZ basis and aug-cc-pVQZ auxiliary basis with respect to full NAF values. The SS-FNO truncation threshold has been kept at $10^{−4}$.}}
    \label{fig:nafthresh}
\end{figure}

Truncation based on occupation number thresholds is more robust than truncation based on a fixed percentage of retained virtual orbitals in natural-orbital-based wave-function methods.~\cite{kallaySystematicWayCost2014}  In occupation-number-based truncation schemes, virtual natural orbitals with occupation numbers above a chosen threshold are retained in the calculations. Fig.~\ref{fig:thresh} shows the convergence of the error in EA values at different natural-orbital truncation thresholds relative to the canonical EA-ADC(3) results. It can be seen that the SS-FNO-based scheme converges more rapidly than the corresponding FNO-based one. The EA values in the SS-FNO-based scheme converge a truncation threshold $10^{-5}$, while the FNO-based scheme shows a large error of -0.641 eV at the same threshold. This demonstrates that the SS-FNO framework is a more suitable approach for reduced-cost EA-ADC(3) calculations. The inclusion of the perturbative correction significantly enhances the accuracy, and the error converges at the threshold of $10^{-4}$. At this threshold, SS-FNO-EA-ADC(3) selects only $38.6\%$ of the total virtual space. 
Further reduction in computational cost can be achieved by reducing the auxiliary basis dimension using Natural Auxiliary Functions (NAFs). Fig.~\ref{fig:nafthresh} illustrates the convergence of the error (in eV) in the EA value of O$_3$ relative to the canonical reference, with decreasing NAF threshold and a fixed SS-FNO threshold of $10^{-4}$ in the EA-ADC(3) method. The EA values (in eV) at different NAF thresholds are listed in Table S3. The aug-cc-pVQZ basis set and the aug-cc-pVQZ/C auxiliary were used for the calculations. The convergence plot suggests that even at the NAF threshold of $10^{-1}$, the EA value converges with an error of 0.022 eV while retaining only about $70\%$ of the auxiliary basis functions. With the inclusion of perturbative corrections, the EA values show a negligible error ($< 0.01$ eV) at the threshold of $10^{-1}$.  However, for further benchmarking of different closed-shell systems, a conservative NAF threshold of $10^{-2}$ was chosen.

%\begin{widetext}
%\FloatBarrier

\subsection{Benchmarking on EA24 test set}
\label{sec:EA24}

%\begin{widetext}
% EA24
\begin{figure*}[htbp]
\centering
    % First row
    \begin{subfigure}{0.16\textwidth}
        \captionsetup{labelformat=empty}
        \includegraphics[width=0.6\linewidth]{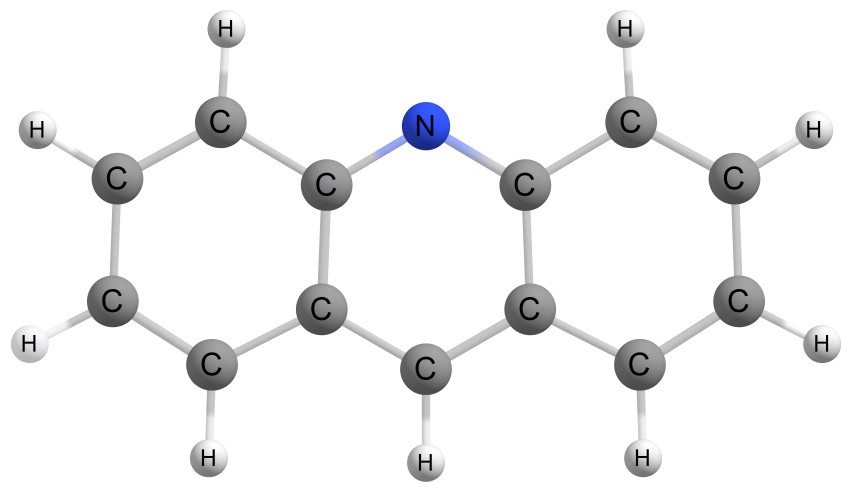}
        \caption{Acridine}
    \end{subfigure}
    %\hspace{0.001\textwidth}
    \begin{subfigure}{0.16\textwidth}
        \captionsetup{labelformat=empty}
        \includegraphics[width=0.6\linewidth]{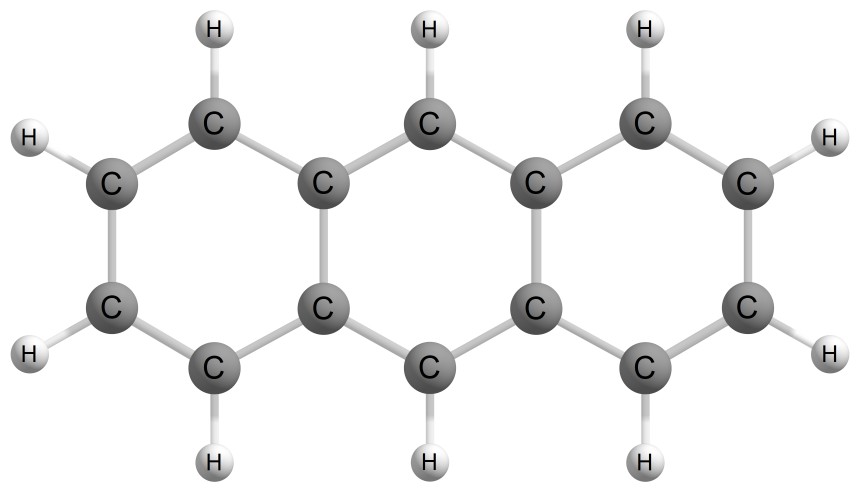}
        \caption{Anthracene}
    \end{subfigure}
    %\hspace{0.001\textwidth}
    \begin{subfigure}{0.16\textwidth}
        \captionsetup{labelformat=empty}
        \includegraphics[width=0.5\linewidth]{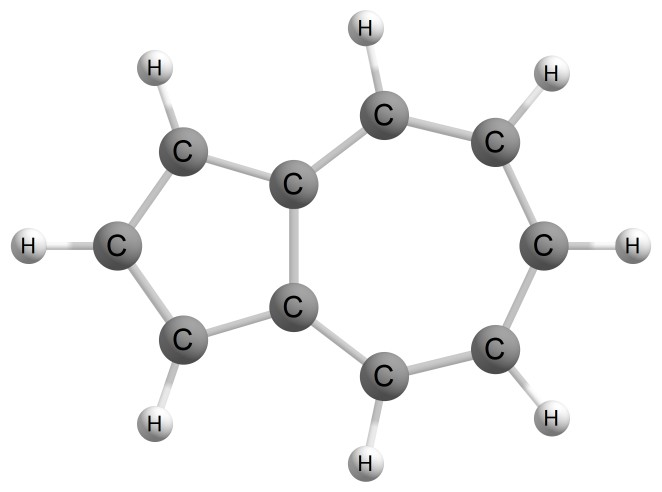}
        \caption{Azulene}
    \end{subfigure}
    %\hspace{0.001\textwidth}
    \begin{subfigure}{0.16\textwidth}
        \captionsetup{labelformat=empty}
        \includegraphics[width=0.5\linewidth]{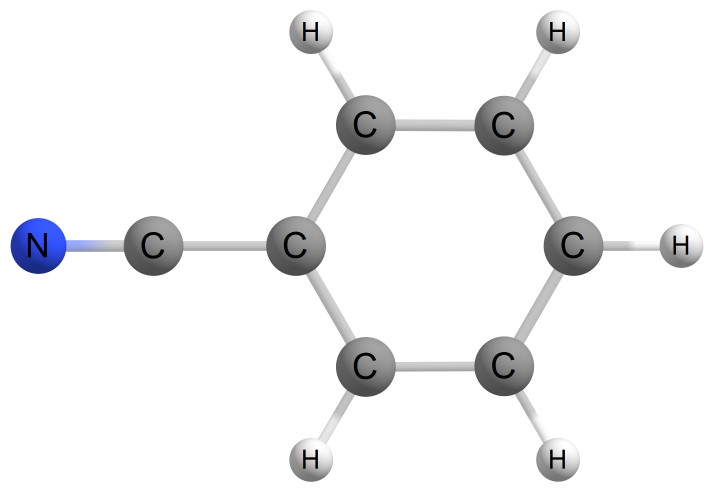}
        \caption{Acridine}
    \end{subfigure}
    %\hspace{0.001\textwidth}
    \begin{subfigure}{0.16\textwidth}
        \captionsetup{labelformat=empty}
        \includegraphics[width=0.5\linewidth]{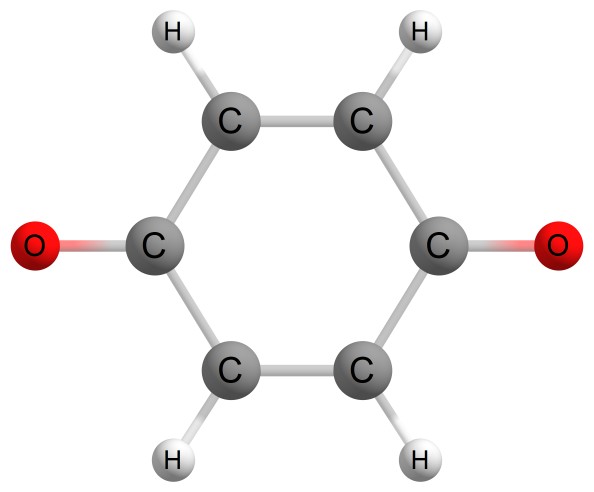}
        \caption{Benzoquinone}
    \end{subfigure}
    %\hspace{0.001\textwidth}
    \begin{subfigure}{0.16\textwidth}
        \captionsetup{labelformat=empty}
        \includegraphics[width=0.6\linewidth]{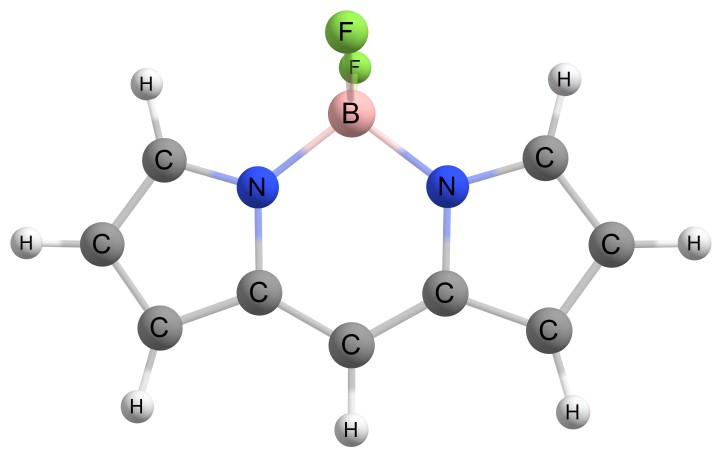}
        \caption{Bodipy}
    \end{subfigure}
    \vspace{0.0025\textwidth}
    
    % Second row
    \begin{subfigure}{0.16\textwidth}
        \captionsetup{labelformat=empty}
        \includegraphics[width=0.6\linewidth]{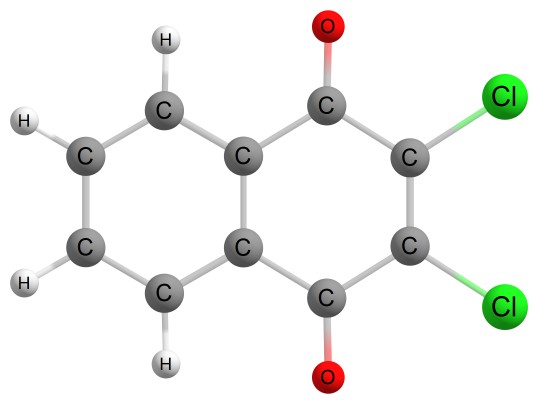}
        \caption{Dichlone}
    \end{subfigure}
    %\hspace{0.025\textwidth}
    \begin{subfigure}{0.16\textwidth}
        \captionsetup{labelformat=empty}
        \includegraphics[width=0.5\linewidth]{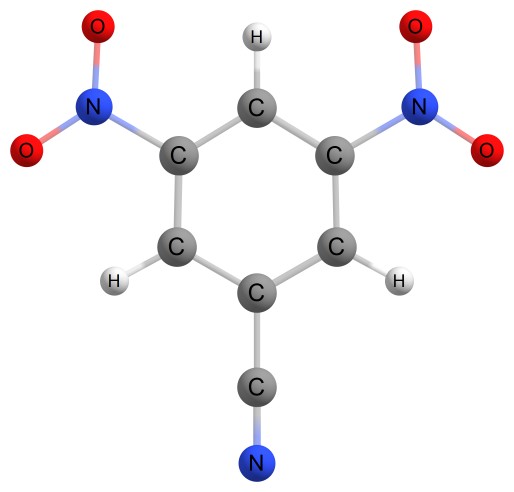}
        \caption{Dinitrobenzonitrile}
    \end{subfigure}
    \begin{subfigure}{0.16\textwidth}
        \captionsetup{labelformat=empty}
        \includegraphics[width=0.5\linewidth]{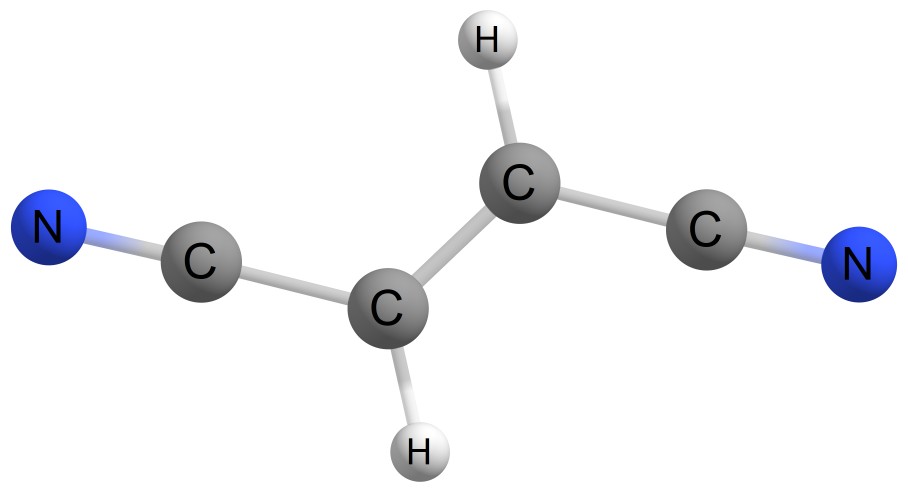}
        \caption{Fumaronitrile}
    \end{subfigure}
    %\hspace{0.025\textwidth}
    \begin{subfigure}{0.16\textwidth}
        \captionsetup{labelformat=empty}
        \includegraphics[width=0.5\linewidth]{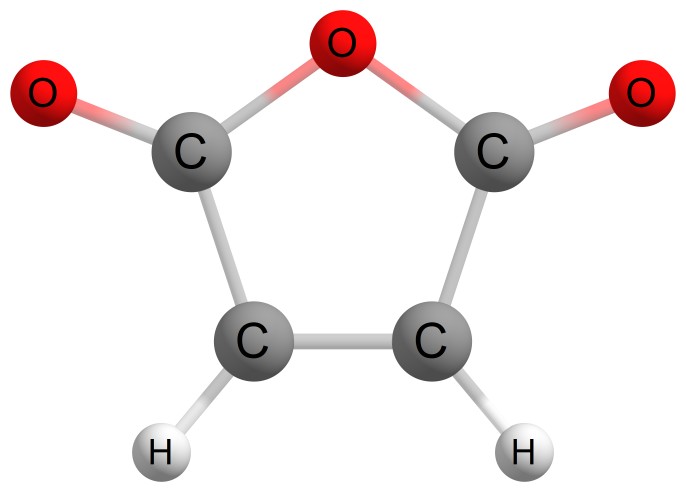}
        \caption{Malaic Anhydride}
    \end{subfigure}
    %\hspace{0.025\textwidth}
    \begin{subfigure}{0.16\textwidth}
        \captionsetup{labelformat=empty}
        \includegraphics[width=0.6\linewidth]{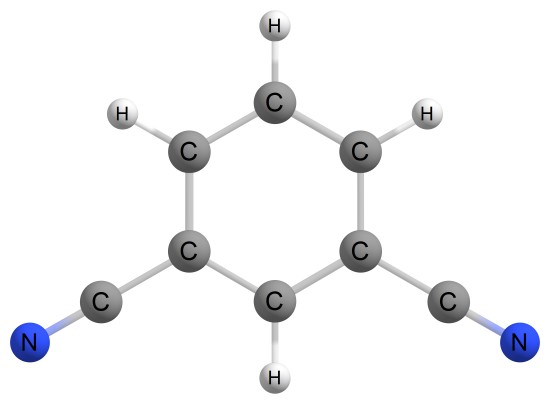}
        \caption{mDCNB}
    \end{subfigure}
    %\hspace{0.025\textwidth}
    \begin{subfigure}{0.16\textwidth}
        \captionsetup{labelformat=empty}
        \includegraphics[width=0.55\linewidth]{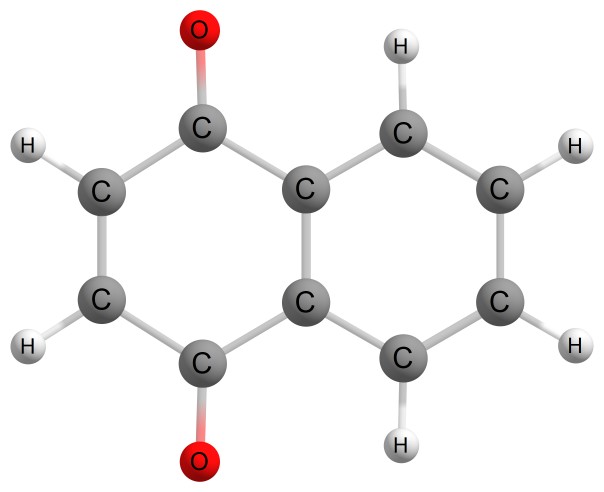}
        \caption{Naphthalenedione}
    \end{subfigure}
    \vspace{0.0025\textwidth}

    % Third row
    \begin{subfigure}{0.16\textwidth}
        \captionsetup{labelformat=empty}
        \includegraphics[width=0.5\linewidth]{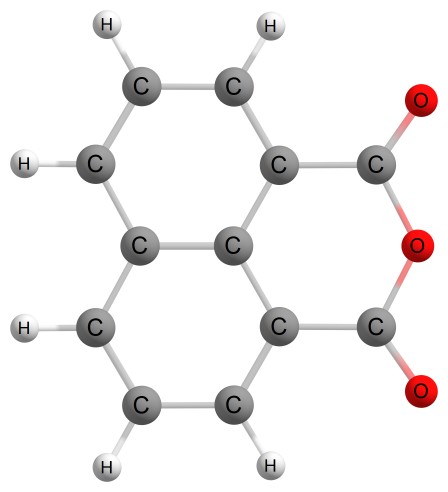}
        \caption{NDCA}
    \end{subfigure}
    %\hspace{0.025\textwidth}
    \begin{subfigure}{0.16\textwidth}
        \captionsetup{labelformat=empty}
        \includegraphics[width=0.65\linewidth]{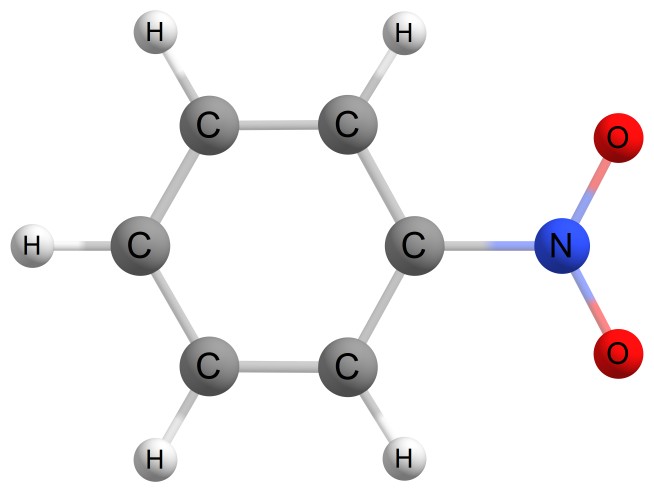}
        \caption{Nitrobenzene}
    \end{subfigure}
    %\hspace{0.025\textwidth}
    \begin{subfigure}{0.16\textwidth}
        \captionsetup{labelformat=empty}
        \includegraphics[width=0.75\linewidth]{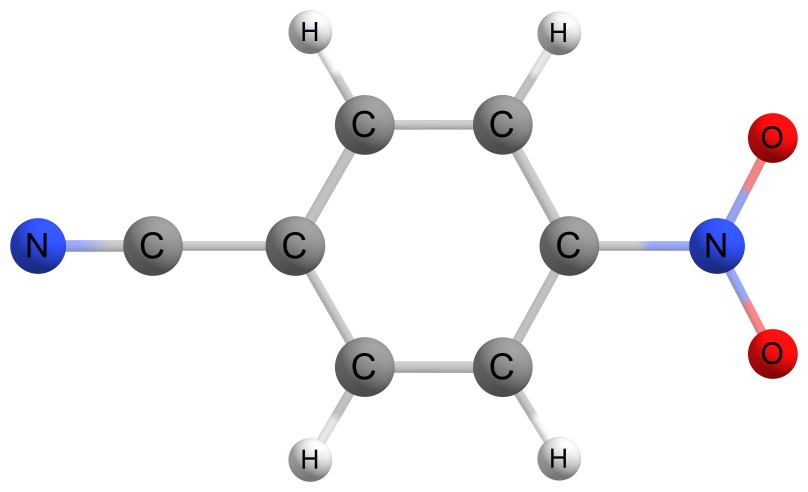}
        \caption{Nitrobenzonitrile}
    \end{subfigure}
    %\hspace{0.025\textwidth}
    \begin{subfigure}{0.16\textwidth}
        \captionsetup{labelformat=empty}
        \includegraphics[width=0.8\linewidth]{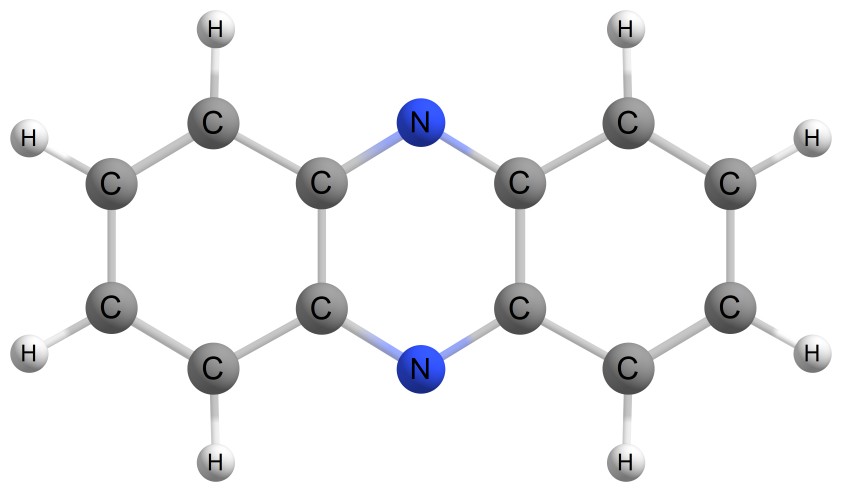}
        \caption{Phenazene}
    \end{subfigure}
    \begin{subfigure}{0.16\textwidth}
        \captionsetup{labelformat=empty}
        \includegraphics[width=0.55\linewidth]{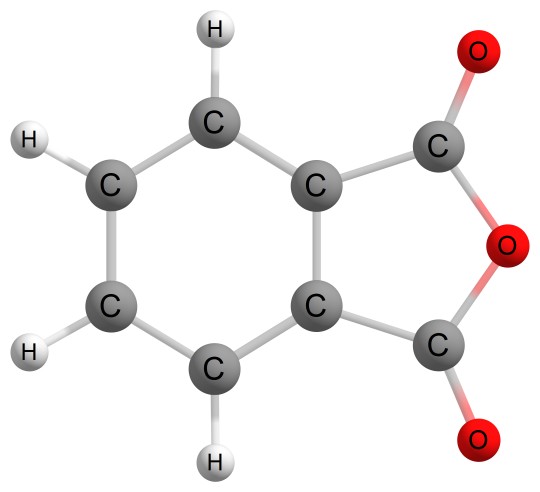}
        \caption{Phthalic Anhydride}
    \end{subfigure}
    %\hspace{0.025\textwidth}
    \begin{subfigure}{0.16\textwidth}
        \captionsetup{labelformat=empty}
        \includegraphics[width=0.65\linewidth]{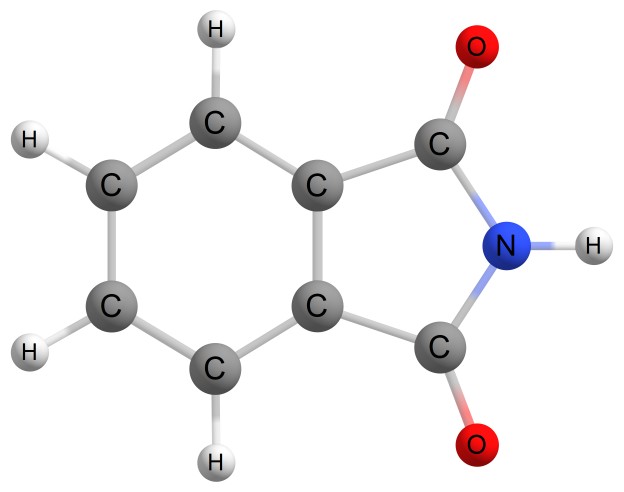}
        \caption{Phthalimide}
    \end{subfigure}
    \vspace{0.0025\textwidth}

    % Fourth row
    \begin{subfigure}{0.16\textwidth}
        \captionsetup{labelformat=empty}
        \includegraphics[width=0.35\linewidth]{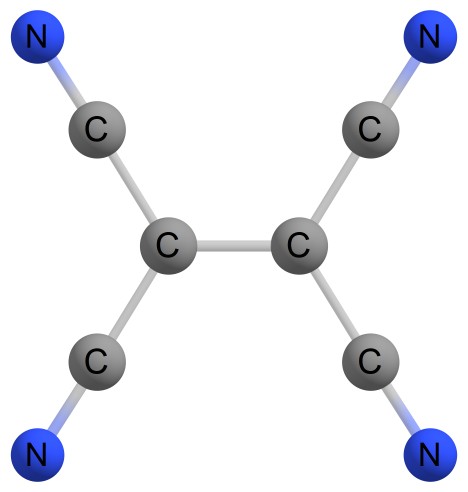}
        \caption{TCNE}
    \end{subfigure}
    %\hspace{0.025\textwidth}
    \begin{subfigure}{0.16\textwidth}
        \captionsetup{labelformat=empty}
        \includegraphics[width=0.65\linewidth]{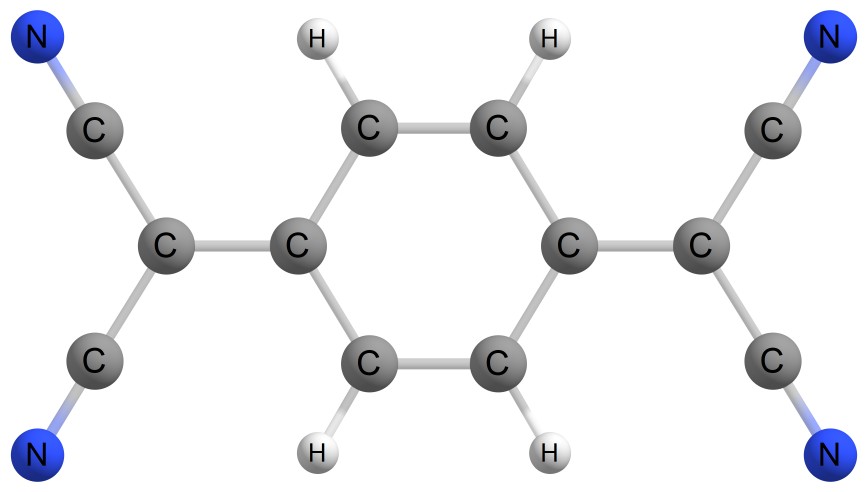}
        \caption{TCNQ}
    \end{subfigure}
    \begin{subfigure}{0.16\textwidth}
        \captionsetup{labelformat=empty}
        \includegraphics[width=0.5\linewidth]{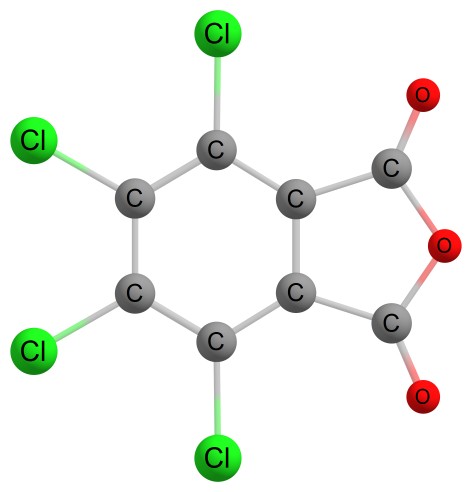}
        \caption{\mbox{Cl$_4$-isobenzofurandione}}
    \end{subfigure}
    %\hspace{0.025\textwidth}
    \begin{subfigure}{0.16\textwidth}
        \captionsetup{labelformat=empty}
        \includegraphics[width=0.5\linewidth]{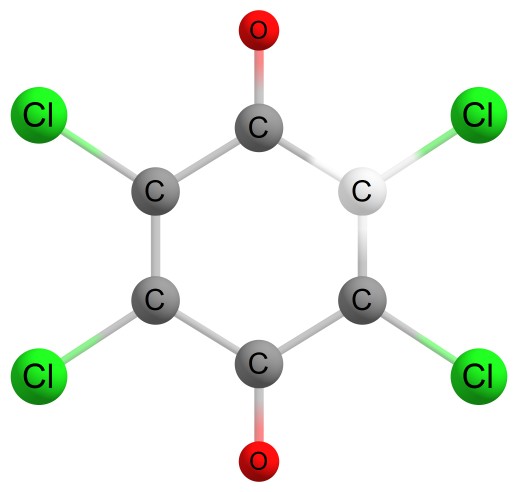}
        \caption{Cl$_4$-benzoquinone}
    \end{subfigure}
    %\hspace{0.025\textwidth}
    \begin{subfigure}{0.16\textwidth}
        \captionsetup{labelformat=empty}
        \includegraphics[width=0.75\linewidth]{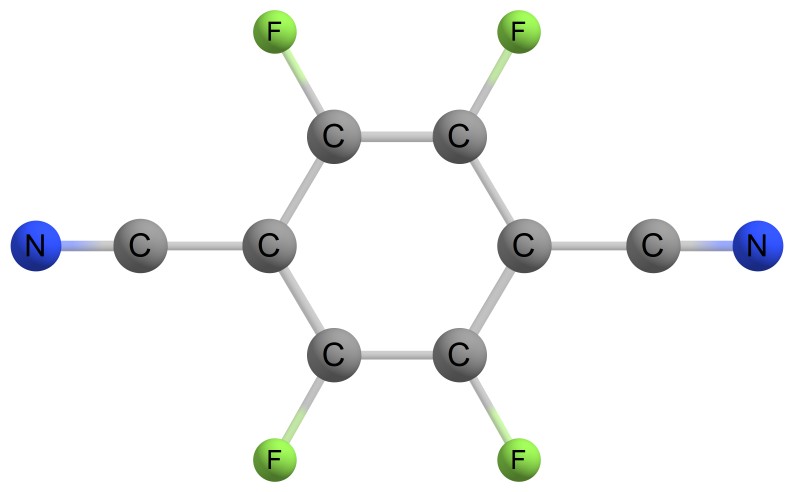}
        \caption{\mbox{F$_4$-benzenedicarbonitrile}}
    \end{subfigure}
    %\hspace{0.052\textwidth}
    \begin{subfigure}{0.16\textwidth}
        \captionsetup{labelformat=empty}
        \includegraphics[width=0.5\linewidth]{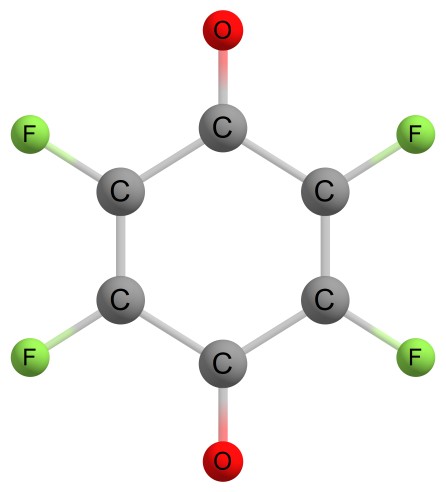}
        \caption{F$_4$-benzoquinone}
    \end{subfigure}
    %\vspace{0.01\textwidth}

    % Main caption for the figure
    \caption{
    \label{fig:EA24}
    The molecules in the EA24 test set.}
\end{figure*}

%\FloatBarrier
%\end{widetext}

To further assess the performance of the SS-FNO-ADC(3) method, we calculated the electron affinities of the molecules included in the EA24 test set of Sherrill and co-workers.~\cite{richardAccurateIonizationPotentials2016} This set consists of 24 organic acceptor molecules that possess bound valence-type electron-attached states, for which experimental values are available for most systems. The electron affinity values of the molecules in EA24 test set have been extensively studied and benchmarked for EA due to its practical implications in organic photovoltaics.~\cite{richardAccurateIonizationPotentials2016,gallandiAccurateIonizationPotentials2016,knightAccurateIonizationPotentials2016,dolgounitchevaCorrectionAccurateIonization2017,duttaDomainbasedLocalPair2019,shaalanalagAccuratePredictionVertical2022,behjouElectronAffinitiesEquationofMotion2025,opokuInitioElectronPropagators2024,opokuNewGenerationElectronPropagatorMethods2024,mesterVerticalIonizationPotentials2023,galynskaExploringElectronAffinities2024} The molecules included in the EA24 test set are shown in Fig. \ref{fig:EA24}, and their geometries were taken from Ref.~\onlinecite{knightAccurateIonizationPotentials2016}. Some molecules possess more than one bound electron-attached state. However, we have considered only the lowest energy electron-attached state for the sake of simplicity. 
% EDP_SSFNO
\begin{figure}[h!]
    \centering
    \includegraphics[width=0.45\textwidth]{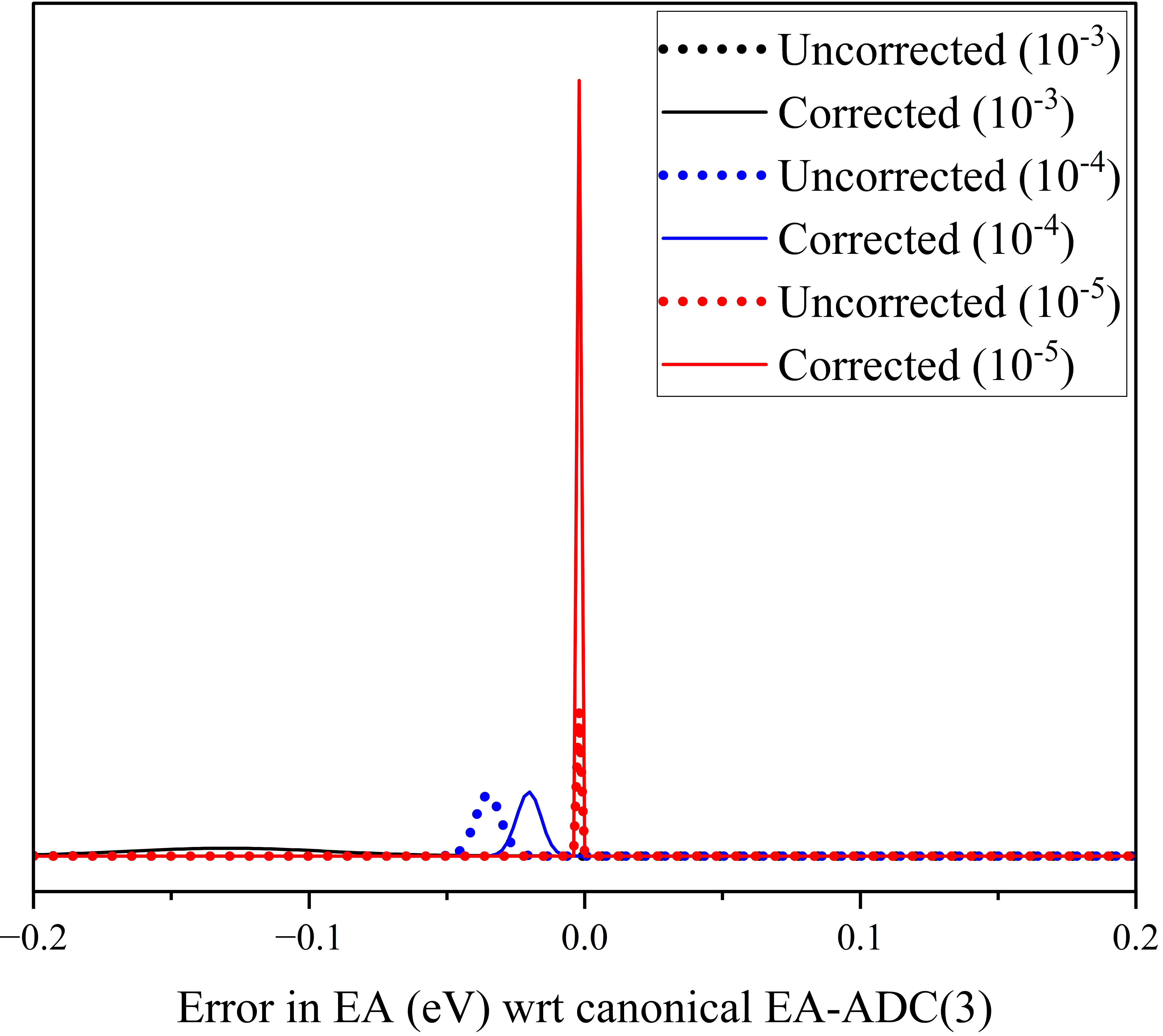}
    \caption{\justifying{The distribution of error in EA using SS-FNO-EA-ADC(3) method in aug-cc-pVDZ basis and aug-cc-pVDZ/C auxiliary basis with respect to the canonical EA-ADC(3) values (in eV) for the EA24 test set at different SS-FNO truncation thresholds.}}
    \label{fig:edp_ssfno}
\end{figure}

The vertical electron-attachment (EA) energies of these molecules were calculated at three truncation thresholds, 
 $10^{-3}$, $10^{-4}$, and $10^{-5}$,using the aug-cc-pVDZ basis set together with the aug-cc-pVDZ/C auxiliary basis set. Fig.~\ref{fig:edp_ssfno} shows the distribution of errors relative to the canonical EA-ADC(3) results. The plot shows a smooth convergence of the EA values as the truncation threshold is tightened. Furthermore, the inclusion of perturbative corrections leads to a noticeable reduction in both the spread and magnitude of the errors. The corresponding statistical parameters are summarized in Table~\ref{tab:edp_ssfno}using the following measures: MAD (maximum absolute deviation), ME (Mean Error), MAE (mean absolute error), STD (standard deviation), and RMSD (root mean squared deviation).The MAD value decreases from 0.378 eV at the $10^{-3}$ threshold to 0.042 eV at $10^{-4}$, and further to 0.003 eV at $10^{-5}$. The inclusion of the perturbative correction substantially improves the results, particularly at higher thresholds. For instance, at the $10^{-3}$ threshold, the MAD reduces from 0.378 eV to 0.170 eV upon applying the perturbative correction, while at $10^{-4}$, it decreases from 0.042 eV to 0.025 eV. At the tightest threshold ($10^{-5}$), the perturbative correction has a negligible effect, indicating convergence with respect to the SS-FNO truncation. The ME values suggest that the SS-FNO truncation systematically underestimates the EA, as reflected by the identical magnitudes of ME and MAE. All statistical parameters become negligible at the $10^{-5}$ threshold, confirming near-complete convergence of the SS-FNO approach.

%\begin{widetext}
%\FloatBarrier
% STAT_EDP_SSFNO table.
\begin{table*}[htbp]
\caption{
\label{tab:edp_ssfno}
\justifying{Statistical analysis of errors (in eV) of SS-FNO-EA-ADC(3) with respect to the canonical results at different FNO thresholds.}
}
\begin{ruledtabular}
\begin{tabular}{ l c c c c c c }
SS-FNO threshold &\multicolumn{2}{c}{$10^{-3}$} &\multicolumn{2}{c}{$10^{-4}$} &\multicolumn{2}{c}{$10^{-5}$} \\
\cline{2-3}
\cline{4-5}
\cline{6-7}
 &Uncorrected &Corrected &Uncorrected &Corrected &Uncorrected &Corrected \\
\hline
MAD    & 0.378 & 0.170 & 0.042 & 0.025 & 0.003 & 0.003 \\
ME     &-0.329 &-0.130 &-0.035 &-0.020 &-0.003 &-0.002 \\
MAE    & 0.329 & 0.130 & 0.035 & 0.020 & 0.003 & 0.002 \\
STD    & 0.056 & 0.036 & 0.005 & 0.004 & 0.000 & 0.000 \\
RMSD   & 0.333 & 0.135 & 0.035 & 0.021 & 0.003 & 0.002 \\
\end{tabular}
\end{ruledtabular}
\end{table*}
%\FloatBarrier
%\end{widetext}

% DOT_SSFNO
\begin{figure}[htbp]
    \centering
    \includegraphics[width=0.45\textwidth]{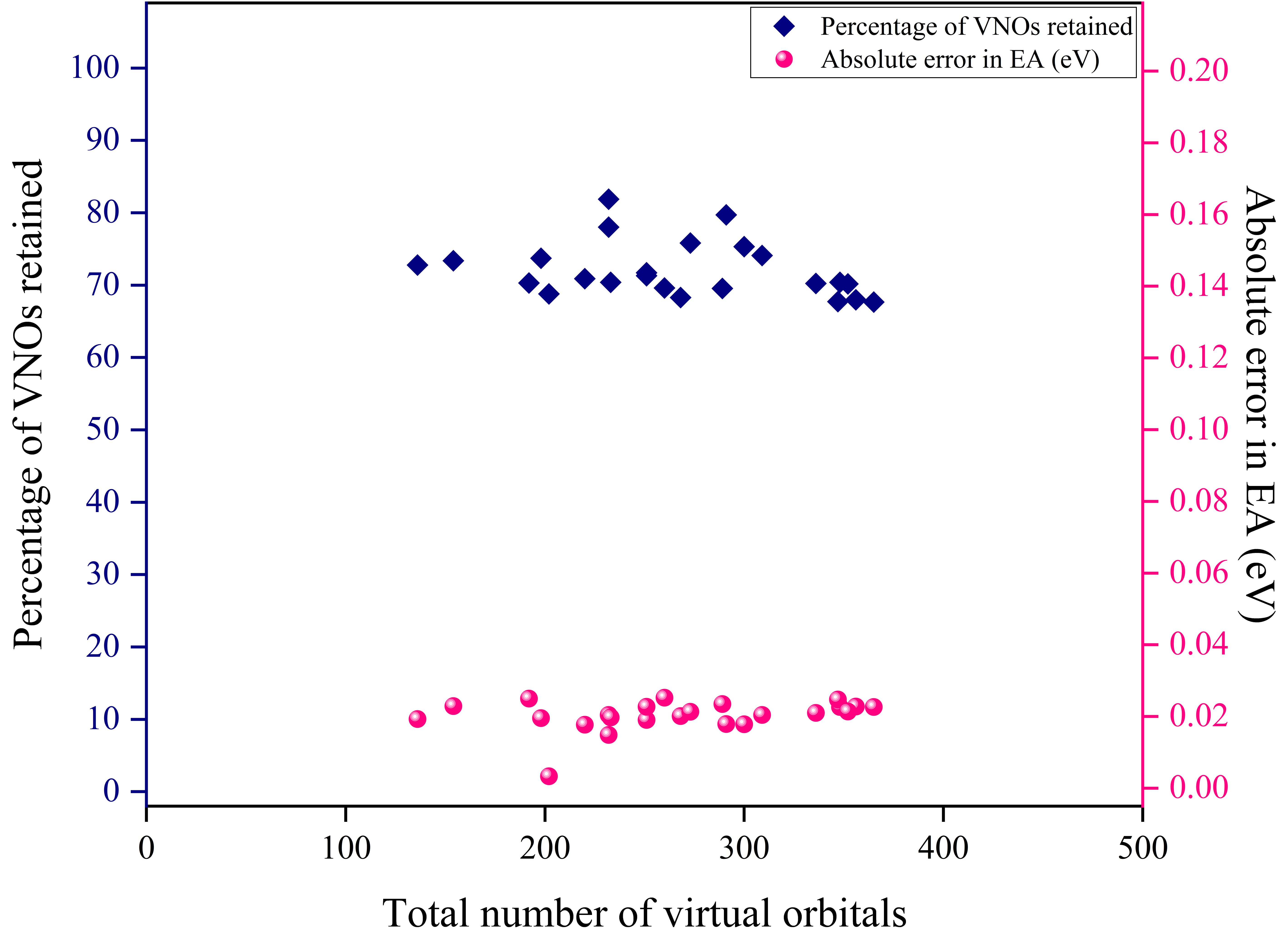}
        \caption{\justifying{Percentage of retained SSFNOs and absolute errors in EA values (eV) relative to canonical results, as a function of system size, for EA24 molecules using the aug-cc-pVDZ basis and aug-cc-pVDZ/C auxiliary basis}}
    \label{fig:dot_ssfno}
\end{figure}
\begin{figure}[htbp]
    \centering
    \includegraphics[width=0.45\textwidth]{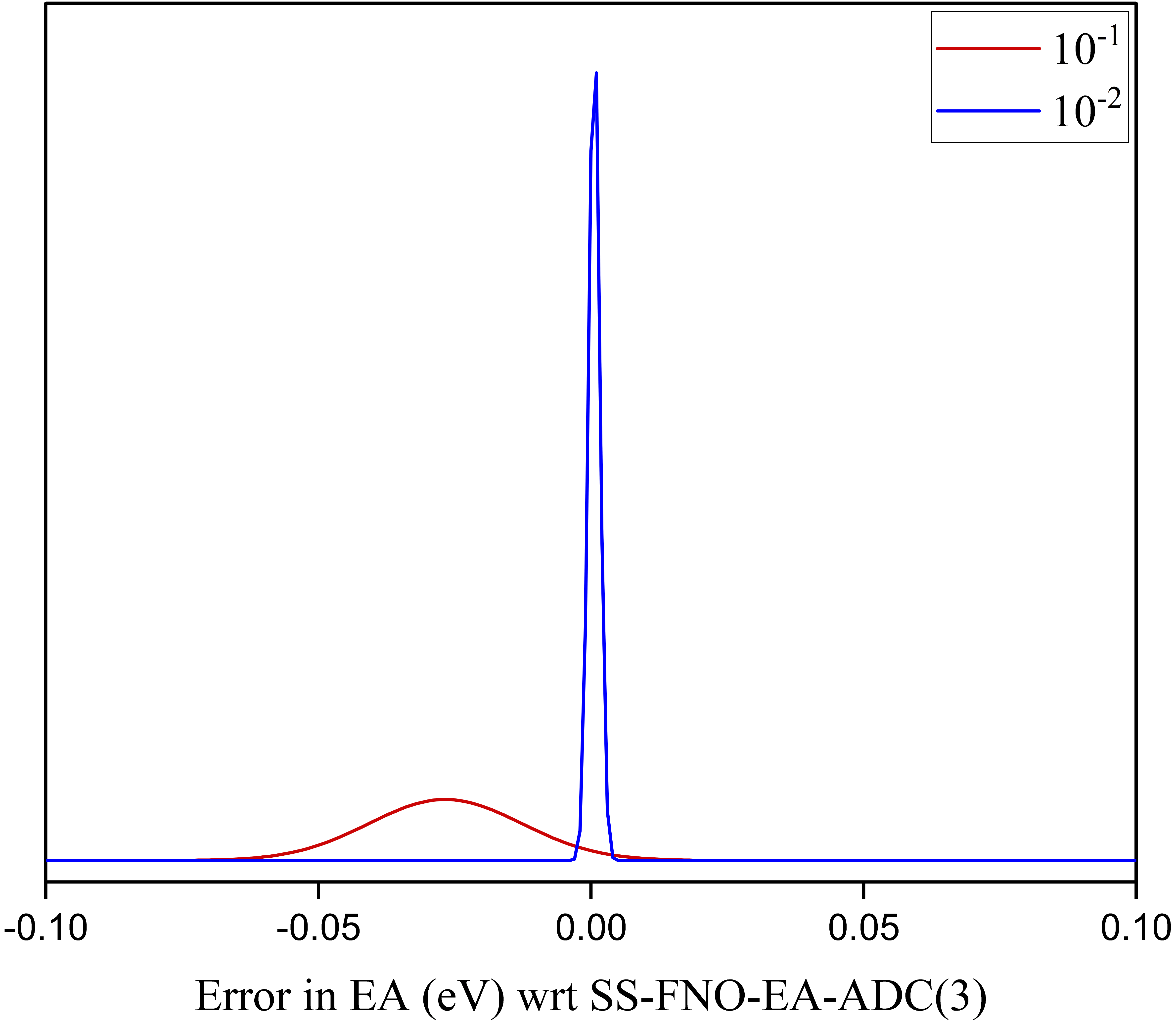}
    \caption{\justifying{The distribution of error of EA values (in eV) using SS-FNO-EA-ADC(3) method in aug-cc-pVDZ basis and aug-cc-pVDZ/C auxiliary basis at truncated SS-FNO threshold ($10^{-4}$) with respect to the full NAF results for truncated natural auxiliary functions.}}
    \label{fig:edp_naf}
\end{figure}

Fig.~\ref{fig:dot_ssfno} shows the percentage of retained SSFNO (blue) and the absolute error in the EA energy (pink) as a function of system size. The percentage of retained SS-FNOs exhibits a variation ranging from approximately 65\% to 85\% across different systems. The truncation introduces a nearly uniform error across the test set, with the maximum absolute deviation remaining below 0.025 eV (including the perturbative correction). It demonstrates that the SS-FNO method can select an appropriate active space largely independent of the system size. The degree of truncation becomes even larger when larger basis sets are employed. For example, the average truncation for the EA24 test set is approximately 52 $\%$ in the aug-cc-pVTZ basis set (See Table S7).
We also performed a statistical analysis of the errors arising from NAF truncation. Fig.~\ref{fig:edp_naf} shows the distribution of errors relative to the canonical results for the SS-FNO-EA-ADC(3) method evaluated at different NAF truncation thresholds ($10^{-1}$ and $10^{-2}$). The perturbative correction is included for this comparison. 
%\newpage
% EDP_NAF

%\begin{widetext}
%\FloatBarrier
% STAT_EDP_NAF table.
\begin{table*}[htbp]
\caption{
\label{tab:edp_naf}
\justifying{Statistical parameters on errors (in eV) of corrected SS-FNO-EA-ADC(3) with respect to full NAF results at different NAF thresholds. The SS-FNO threshold is kept at $10^{-4}$.}
}
\begin{ruledtabular}
\begin{tabular}{ l c c }
 &$10^{-1}$ &$10^{-2}$ \\
\hline
MAD    & 0.048 & 0.003 \\
ME     &-0.027 & 0.001 \\
MAE    & 0.027 & 0.001 \\
STD    & 0.014 & 0.001 \\
RMSD   & 0.030 & 0.001 \\
\end{tabular}
\end{ruledtabular}
\end{table*}
%\FloatBarrier
%\end{widetext}
Table S5 in the Supplementary Material reports the vertical electron attachment energies (in eV) for all molecules in the EA24 test set, computed using the SS-FNO-EA-ADC(3) method with the SS-FNO truncation threshold fixed at $10^{-4}$ and the NAF thresholds set to $10^{-1}$ and $10^{-2}$. The statistical analysis presented in Table \ref{tab:edp_naf} reveals that, at a NAF threshold of $10^{-2}$, the deviations in the calculated EA values are negligible, whereas a threshold of $10^{-1}$ results in a mean absolute deviation (MAD) of only 0.045 eV. These findings indicate that the adoption of a NAF threshold of $10^{-2}$ ensures a reliable and accurate description of electron attachment energies within the SS-FNO-EA-ADC(3) framework. 

\begin{figure}[htbp]
    \centering
    \includegraphics[width=0.45\textwidth]{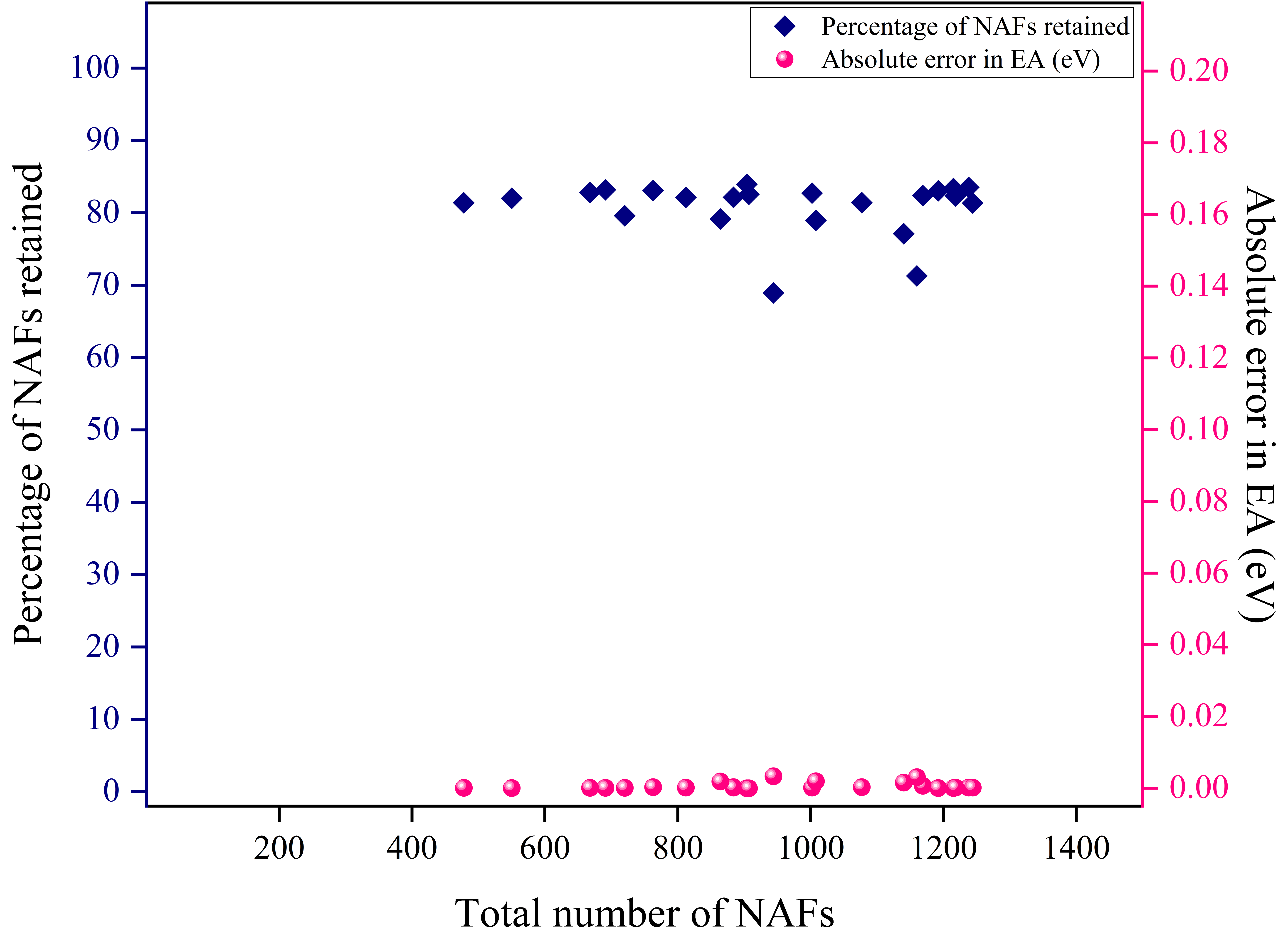}
    \caption{\justifying{The percentage of the NAFs retained (NAF threshold $10^{-2}$) and absolute error in EA values (eV) with respect to results with full NAF as the size of the auxiliary basis for EA24 test set molecules in aug-cc-pVDZ basis and aug-cc-pVDZ/C auxiliary basis.}}
    \label{fig:dot_naf}
\end{figure}

Fig.~\ref{fig:dot_naf} illustrates the percentage of retained NAFs across different systems in the EA24 test set, together with the corresponding truncation errors. At a NAF threshold of $10^{-2}$, the proportion of retained NAFs ranges from approximately 70 $\%$ to 90$\%$ across the test set. The errors introduced by NAF truncation, relative to the full NAF reference, exhibit a highly systematic behavior and remain below 0.01 eV for all systems, indicating excellent numerical stability. In contrast, employing a more aggressive NAF threshold of $10^{-1}$ significantly reduces the number of retained NAFs to approximately 20–30$\%$ across all systems (see Fig. S1). However, the associated deviations are comparatively larger in magnitude than those arising from the truncation of the natural orbital space.
 % DOT_NAF

From the results of the benchmark study on the EA24, it is evident that the SS-FNO threshold of $10^{-4}$ with NAF truncation at $10^{-2}$ threshold provides an optimal balance between computational cost and accuracy. These thresholds are therefore adopted for all subsequent calculations presented in this manuscript.

\begin{table*}[htbp]
\caption{
\label{tab:ea24_ccsd(t)}
\justifying{The electron affinities (in eV) of the first electron attached states of EA24 test set molecules calculated using various EA-ADC methods in CBS limit (in the SS-FNO basis, including perturbative correction) and compared with CCSD(T) results. The EA-EOM-DLPNO-CCSD/CBS results are also shown for comparison. The experimental values are provided for reference. The SS-FNO threshold is kept at $10^{-4}$ and NAF threshold at $10^{-2}$.}
}
\resizebox{\textwidth}{!}{
\begin{ruledtabular}
\begin{tabular}{ l c c c c c c }
Molecule &ADC(2)\footnotemark[1] &SS-FNO-ADC(3) &SS-FNO-sm-ADC[(2)+x(3)] &EOM-DLPNO-CCSD\footnotemark[2] &CCSD(T)~\cite{richardAccurateIonizationPotentials2016} &Expt.~\cite{informaticsNISTChemistryWebBook} \\
\hline
Acridine	                           &1.26	&0.46	&0.79	&0.65  &0.69	&0.9  \\
Anthracene	                           &0.93	&0.10	&0.45	&0.31  &0.28	&0.53 \\
Azulene	                               &1.19	&0.31	&0.68	&0.59  &0.54	&0.8  \\
Benzonitrile	                       &-0.20	&-0.30	&-0.26	&-0.35 &-0.21	&0.26 \\
Benzoquinone	                       &2.12	&1.31	&1.66	&1.50  &1.71	&1.85 \\
Bodipy	                               &2.22	&1.51	&1.79	&1.70  &1.67	&     \\
Dichlone	                           &2.66	&1.48	&1.99	&1.83  &1.92	&2.21 \\
Dinitrobenzonitrile	                   &2.57	&1.52	&1.96	&1.82  &1.76	&2.16 \\
Fumaronitrile	                       &1.64	&0.76	&1.14   &1.02  &0.98	&1.25 \\
Maleic anhydride	                   &1.62	&0.70	&1.11	&1.00  &1.01	&1.44 \\
mDCNB	                               &1.30	&0.36	&0.76	&0.65  &0.61	&0.91 \\
NDCA	                               &1.90	&0.96	&1.37	&1.26  &1.26	&     \\
Naphthalenedione	                   &2.07	&1.17	&1.56	&1.38  &1.47	&1.81 \\
Nitrobenzene	                       &1.11	&0.36	&0.66	&0.52  &0.54	&1    \\
Nitrobenzonitrile	                   &1.98	&1.12	&1.48	&1.36  &1.3  	&1.69 \\
Phenazine	                           &1.68	&0.89	&1.22	&1.07  &1.11	&1.31 \\
Phthalic anhydride	                   &1.50	&0.52	&0.95	&0.85  &0.87	&1.25 \\
Phthalimide	                           &1.24	&0.29	&0.70	&0.58  &0.63	&1.02 \\
TCNE	                               &4.02	&2.71   &3.26	&3.20  &3.05	&3.16 \\
TCNQ	                               &4.28	&3.07	&3.59	&3.48  &3.33	&2.8  \\
Cl$_4$-isobenzofuranedione	           &2.53	&1.13	&1.74	&1.66  &1.68	&1.96 \\
Cl$_4$-benzoquinone	                   &3.30	&1.98	&2.56	&2.41  &2.48	&2.78 \\
F$_4$-benzenedicarbonitrile	           &2.40	&1.26	&1.76	&1.72  &1.62	&1.89 \\
F$_4$-benzoquinone	                   &2.98	&1.93	&2.40	&2.29  &2.29	&2.7  \\
\hline
MAD                                    &0.97   &0.56  &0.26  &0.21  &       &     \\
ME                                     &0.65   &-0.29 &0.11  &-0.01 &       &     \\
MAE                                    &0.65   &0.29  &0.12  &0.06  &       &     \\
STD                                    &0.19   &0.11  &0.07  &0.08  &       &     \\
RMSD                                   &0.68   &0.31  &0.13  &0.08  &       &     \\
\end{tabular}
\end{ruledtabular}
}
\footnotetext[1]{\mbox{in the canonical basis.}}
\footnotetext[1]{\mbox{NormalPNO with TCutPNOSingles $=$ 1e-9 is used.}}
\end{table*}

Table~\ref{tab:ea24_ccsd(t)} summarizes the electron affinities (EAs) of the lowest electron-attached states for all molecules in the EA24 test set, computed using various EA-ADC methods and compared with the CCSD(T) reference values extrapolated to the complete basis set (CBS) limit.~\cite{richardAccurateIonizationPotentials2016} The CBS extrapolation was performed using results obtained with the aug-cc-pVXZ (X = D, T) basis sets. For the second-order ADC calculations, canonical results are reported, whereas for the third-order ADC, only the corrected SS-FNO values are considered for comparison. The ADC(2) method exhibits the largest deviation from the CCSD(T) benchmarks, with a mean absolute deviation (MAD) of 0.973 eV. Inclusion of third-order corrections substantially improves the accuracy, as reflected in the reduced MAD of 0.56 eV for the SS-FNO-ADC(3) method, although this value remains larger than that observed for the corresponding EOM-DLPNO-CCSD method. The SS-FNO-sm-ADC[(2)+x(3)] approach demonstrates significantly improved performance, yielding a MAD of only 0.26 eV and showing close agreement with the EOM-DLPNO-CCSD results, which exhibit slightly smaller MAD and MAE but a larger standard deviation (STD). Although a direct comparison with experimental data is less meaningful due to the omission of vibrational contributions in the calculations, the SS-FNO-sm-ADC[(2)+x(3)] method at the CBS limit nonetheless exhibits marginally better agreement with experiment than the corresponding CCSD(T) results.
\subsection{EA of Non-valence correlation-bound Anions}
\label{sec:C6F6}

% C6F6 ADC table.
\begin{table*}[htbp]
\caption{
\label{tab:C6F6_adc}
\justifying{The electron affinities (in eV) of the non-valence correlation-bound state of C$_6$F$_6$ molecule calculated using various EA-ADC methods and compared with EOM results.}
}
\begin{ruledtabular}
\begin{tabular}{ l c c c c c c c}
SS-FNO threshold &\multicolumn{2}{c}{$10^{-4}$} &\multicolumn{2}{c}{$10^{-5}$} &\multicolumn{2}{c}{$10^{-6}$} &Canonical \\
\cline{2-3}
\cline{4-5}
\cline{6-7}
 &Uncorrected &Corrected &Uncorrected &Corrected &Uncorrected &Corrected \\
\hline

EA-ADC(3)            &-0.225 &-0.157 &-0.109 &-0.105 &-0.074 &-0.075 & 0.021 \\
sm-EA-ADC[(2)+x(3)]     & 0.036 & 0.104 & 0.118 & 0.122 & 0.132 & 0.131 & 0.153 \\

EA-ADC(2)            & &       &  &       &  &       & 0.515 \\
EA-EOM-DLPNO-CCSD & &       &  &       &  &       & 0.039  \\
EA-EOM-MP2~\cite{vooraNonvalenceCorrelationBoundAnion2014}           &       &       &       &       &       &       & 0.135 \\
EA-EOM-CCSD          &       &       &       &       &       &       & 0.133 \\
\end{tabular}
\end{ruledtabular}
\end{table*}

%\FloatBarrier
%\end{widetext}

The additional electron in the non-valence correlation-bound $($NVCB$)$ state is weakly bound in a diffuse orbital through electron correlation. Due to their sensitivity, NVCB anions provide a challenging test case for natural-orbital-based approximate wave-function methods. Several molecules containing NVCB anionic states have been extensively studied in the literature, for example  TCNE,~\cite{vooraTheoreticalApproachesTreating2017} polycyclic aromatic hydrocarbons,~\cite{vooraNonvalenceCorrelationBoundAnion2015} perfluorobenzene,~\cite{vooraNonvalenceCorrelationBoundAnion2014} fullerenes~\cite{vooraNonvalenceCorrelationBoundAnion2014a} etc. IIn this work, we study the NVCB anionic state of perfluorobenzene (C$_6$F$_6$).

Perfluorobenzene possesses a diffuse non-valence correlation-bound anionic state at its planar equilibrium geometry. This state evolves barrierlessly into a valence-bound anion as the molecule buckles to a $C_{2v}$ geometry and may contribute to the high electron mobility observed in liquid C$_6$F$_6$.~\cite{vooraNonvalenceCorrelationBoundAnion2014} The geometry of C$_6$F$_6$ was optimized using ORCA~\cite{neeseSoftwareUpdateORCA2022} using RI-MP2/aug-cc-pVDZ. at the RI-MP2/aug-cc-pVDZ level of theory.

The electron affinity corresponding to the NVCB anion was calculated using the aug-cc-pVTZ basis set with additional $7s7p$ diffuse functions placed on a ghost atom at the center of the molecule. The results are presented in Table~\ref{tab:C6F6_adc}. The EA-ADC(2) method significantly overestimates the electron affinity compared with the EOM-CCSD and EOM-MP2 results. In contrast, the EA-ADC(3) method underestimates the electron affinity. The convergence of EA values is much slower for the NVCB state as compared to the valence states in both ADC(3) and sm-EA-ADC[(2)+x(3)] as the additional electron is distributed over the entire molecule in the former. At a truncation threshold of $10^{-4}$, the sm-EA-ADC[(2)+x(3)] method shows an error of approximately 0.05 eV relative to the canonical result. The error decreases to approximately 0.03 eV when the threshold is tightened to $10^{-5}$. Even at a threshold of $10^{-6}$, the error does not fully converge but remains small at approximately 0.02 eV.

% C6F6_Dyson
\begin{figure}[h!]
    \centering
    \includegraphics[width=\columnwidth]{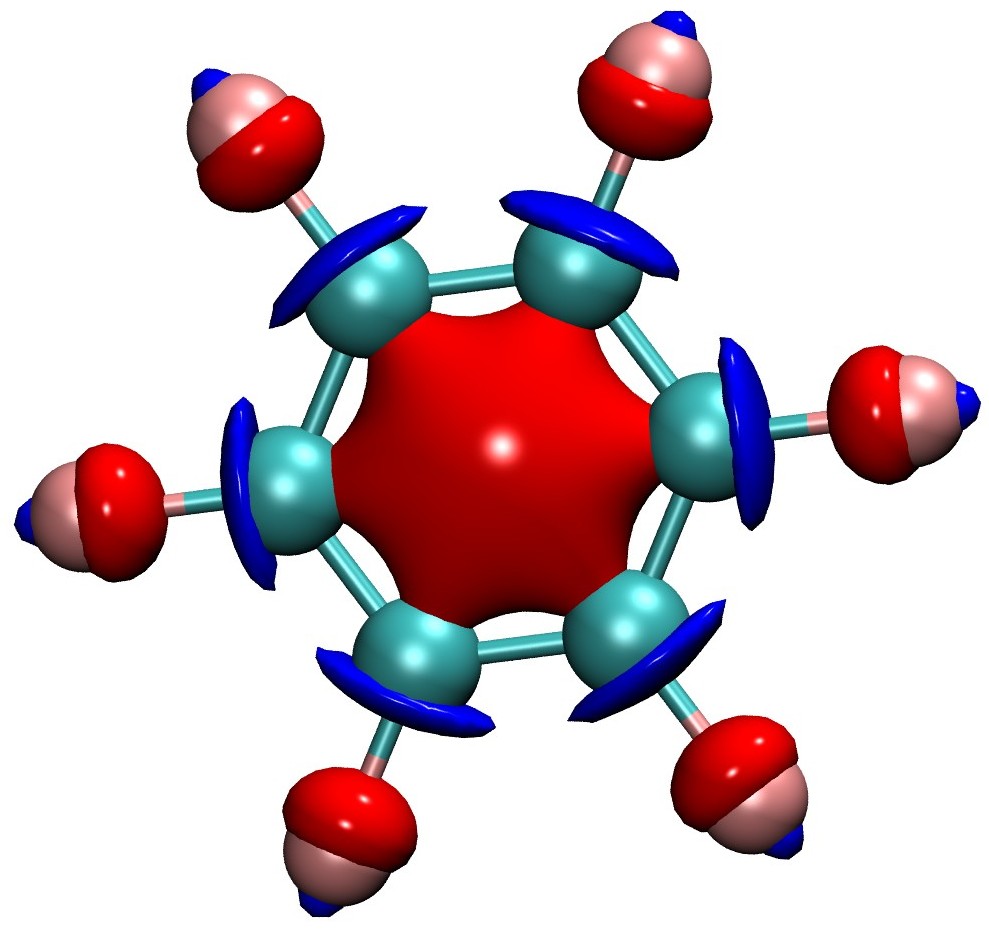}
    \caption{\justifying{The Dyson orbital calculated at the EA-ADC(2) level of theory corresponding to the non-valence correlation-bound EA state of C$_6$F$_6$.}}
    \label{fig:C6F6_Dyson}
\end{figure}

To assess the performance of the SS-FNO truncation (including perturbative correction) within the sm-EA-ADC[(2)+x(3)] framework, a comparative analysis was carried out against the DLPNO truncation in the EA-EOM-DLPNO-CCSD method for the NVCB state. The DLPNO calculations were performed using the aug-cc-pVTZ basis set in conjunction with the aug-cc-pVTZ/C auxiliary basis set. At the canonical level, both EA-EOM-CCSD and sm-EA-ADC[(2)+x(3)] yield comparable electron attachment energies. However, upon truncation, the SS-FNO approach maintains excellent accuracy, exhibiting negligible deviation from the canonical result, whereas the DLPNO method introduces a significant error of 0.094 eV. Even with the \textit{TightPNO} setting, no substantial improvement is observed (see Table S8). This discrepancy can be attributed to the nature of the NVCB anionic state, in which the attached electron is highly delocalized over the entire molecular framework (see Fig.~\ref{fig:C6F6_Dyson}). The EOM-DLPNO-CCSD method, being inherently based on an orbital localization scheme, is unable to adequately describe the delocalized character of the electronic state.
\subsection{Computational Efficiency}
\label{sec:time_calc}
% Zn_Dyson
\begin{figure}[h!]
    \centering
    \includegraphics[width=\columnwidth]{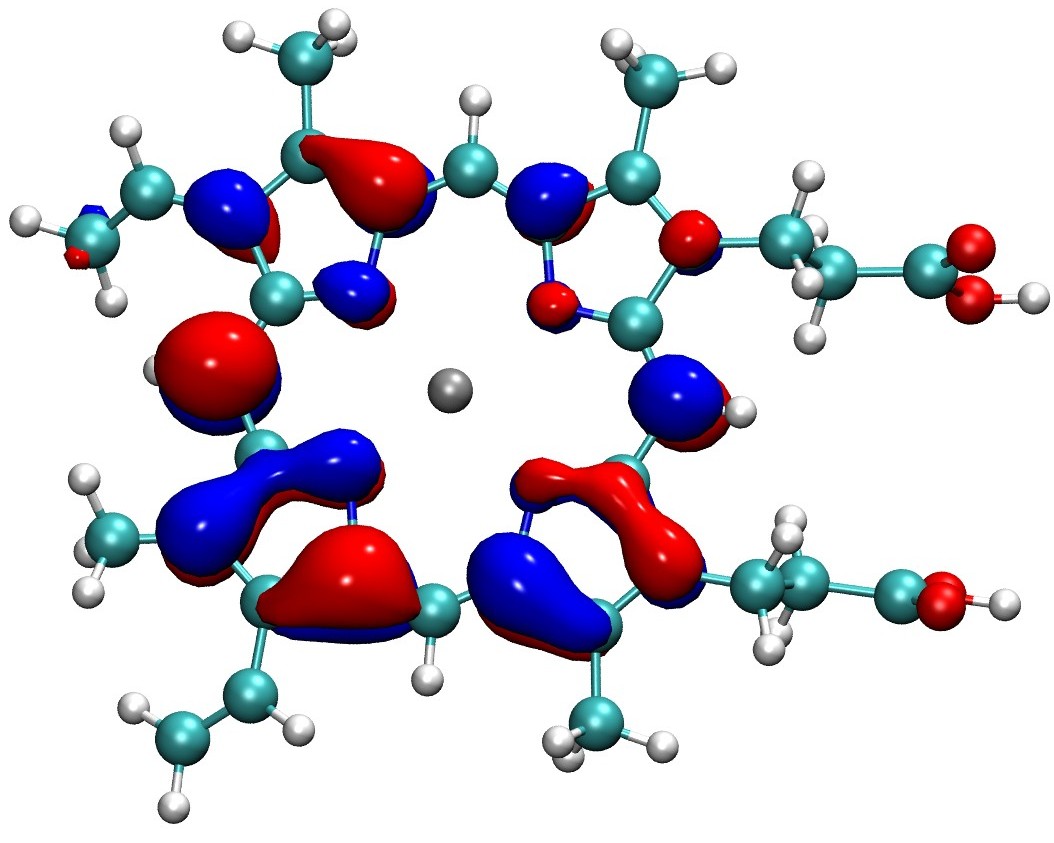}
    \caption{\justifying{The Dyson orbital calculated at the EA-ADC(2) level of theory corresponding to the first EA state of Zn-protoporphyrin.}}
    \label{fig:Zn_Dyson}
\end{figure}

The computational efficiency of the SS-FNO-EA-ADC(3) method was further evaluated by calculating the electron affinity of a large molecular system, Zn-protoporphyrin, which has previously been studied in the context of electron attachment.~\cite{duttaPairNaturalOrbital2016} The molecular geometry was taken from the Supplementary Material of Ref.~\onlinecite{duttaPairNaturalOrbital2016}. The calculation was performed on a dedicated workstation equipped with an Intel Xeon E5-2667 v4 processor (3.20~GHz) and 512~GB of RAM. The aug-cc-pVTZ basis set together with the aug-cc-pVTZ/C auxiliary basis was employed for the Zn atom, while the aug-cc-pVDZ basis set with the aug-cc-pVDZ/C auxiliary basis was used for all other atoms.  Zn-protoporphyrin consists of 75 atoms and 326 electrons, with a total of 1184 virtual orbitals. Applying an SS-FNO truncation threshold of $10^{-4}$ reduced the virtual orbital space to 807 orbitals. 
Using an NAF threshold of $10^{-2}$ further reduced the number of auxiliary functions from 4053 to 3440.  The total computation time for the SS-FNO-EA-ADC(3) calculation was 1 day, 10 hours, and 13 minutes, with the SCF step requiring only 18 minutes. The canonical ADC(2) calculation for a single root required 2 hours and 8 minutes, of which 34 minutes were spent on intermediate generation. In the truncated basis, the ADC(3) calculation took 1 day, 7 hours, and 20 minutes, with 17 hours and 49 minutes required for intermediate construction. The ADC(2) calculation took 20 minutes in the SS-FNO basis. The SS-FNO-EA-ADC(3) method yielded an electron affinity of 1.041~eV for the lowest electron-attached state. The corresponding Dyson orbital for Zn-protoporphyrin is depicted in Fig.~\ref{fig:Zn_Dyson}.

%%%%%%%%%%%%%%%%%%%%%%%%%%%%%%%%%%%%%%%%%%%%%%%%%% Conclusion %%%%%%%%%%%%%%%%%%%%%%%%%%%%%%%%%%%%%%%%%%%%%%%%%%%%%%%%%
\section{Conclusion}
\label{sec5}
We have developed, implemented, and benchmarked a reduced-cost ADC(3) approach for computing electron affinities based on the state-specific frozen natural orbital (SS-FNO) framework within the non-Dyson intermediate-state representation (ISR) formalism. Unlike the conventional MP2-based FNO scheme, the state-specific FNO approach selects an optimal subset of virtual natural orbitals tailored for each electron-attached state, thereby achieving a substantial reduction in computational cost for ADC(3) calculations. To further enhance efficiency, the density-fitting approximation was employed to eliminate the need for storing four-index electron-repulsion integrals. Additional truncation of the auxiliary basis using natural auxiliary functions (NAFs) provides further computational savings.

The accuracy of the SS-FNO-EA-ADC(3) method can be systematically controlled through two thresholds. An FNO threshold of $10^{-4}$ and a NAF threshold of $10^{-2}$ were found to offer an optimal balance between computational efficiency and accuracy. The inclusion of perturbative corrections for FNO truncation plays a crucial role in improving the quantitative reliability of the results. Although the EA-ADC(3) method substantially improves upon the EA-ADC(2) results, its accuracy remains slightly inferior to that of the EOM-CCSD method. In contrast, the sm-EA-ADC[(2)+x(3)] variant demonstrates performance comparable to the EOM-DLPNO-CCSD approach and accurately reproduces non-valence correlation-bound (NVCB) anionic states, where the EOM-DLPNO-CCSD method exhibits significant errors.

The SS-FNO-EA-ADC(3) framework provides a scalable and accurate alternative for studying electron-attachment processes in large molecular systems. Its extension to systems containing heavy elements through the incorporation of relativistic effects is currently under development.

\section{Supplementary Material}
\label{sec6}
The Supplementary Material contains the programmable expressions for SS-FNO-EA-ADC(3), The EA values of O$_3$ molecule across different truncations, the EA values of EA24 test set molecules in different truncation thresholds and in aug-cc-pVXZ (X = D, T) basis sets, the scatter plot at NAF threshold $10^{-1}$, and the optimized geometry and comparison of SS-FNO and DLPNO schemes of perfluorobenzene molecule.

\begin{acknowledgments}
This work has been supported by the ANRF India((Project No. CRG/2023/002558). AKD acknowledges the research fellowship funded by the EU NextGenerationEU through the Recovery and Resilience Plan for Slovakia under project No. 09I03-03-V04-00117.
\end{acknowledgments}

\section{References}
\label{sec6}
\bibliography{libbbt}

\end{document}